\documentclass[prb,twocolumn,aps,amsmath,amssymb]{revtex4}
\usepackage[T1]{fontenc}
\usepackage[utf8]{inputenc}
\usepackage{fancyhdr}
\usepackage{graphicx}
\usepackage{amsmath}
\usepackage{amssymb}
\usepackage{dcolumn}
\usepackage{float}
\usepackage{bm}
\usepackage{slashbox}
\usepackage{amsfonts,times}
\usepackage{parskip}
\usepackage{appendix}
\usepackage[breaklinks=true,colorlinks,citecolor=blue,linkcolor=blue,urlcolor=blue]{hyperref}

\DeclareMathAlphabet{\bi}{OML}{cmm}{b}{it}
\def\be{\begin{equation}}
\def\ee{\end{equation}}
\def\bearr{\begin{eqnarray}}
\def\eearr{\end{eqnarray}}
\def\la{\langle}
\def\ra{\rangle}

\def\bs{\boldsymbol}
\begin{document}

\title{Orbital magnetization senses the topological phase transition in a spin-orbit coupled $\alpha$-$T_3$ system}
\author{Lakpa Tamang$\color{blue}^1$}
\author{Sonu Verma$\color{blue}^2$}
\author{Tutul Biswas$\color{blue}^1$}
\email{tbiswas@nbu.ac.in}
\normalsize
\affiliation
{$\color{blue}^1$ Department of Physics, University of North Bengal, Raja Rammohunpur-734013, India\\
$\color{blue}^2$ Center for Theoretical Physics of Complex Systems,
Institute for Basic Science (IBS), Daejeon, 34126, Republic of Korea}
\date{\today}
\begin{abstract}
The $\alpha$-$T_3$ system undergoes a topological phase transition(TPT) between two distinct quantum spin-Hall phases across $\alpha=0.5$ when the spin-orbit interaction of Kane-Mele type is taken into consideration.
As a hallmark of such a TPT, we find that the Berry curvature and the orbital magnetic moment change their respective signs across the TPT.
We also find the trails of the TPT in another physical observable, namely, the orbital magnetization(OM) that can be, in principle, detected experimentally through the circular dichroism associated with optical absorption. The topological features of the OM are understood in terms of valley and spin physics.
The valley-resolved OM(VROM) and the spin-resolved OM(SROM) exhibit interesting characteristics related to the valley and the spin Chern number when the chemical potential is tuned in the forbidden gap(s) of the energy spectrum. In particular, we find that the slope of the VROM versus the chemical potential in the forbidden gap changes its sign abruptly across the TPT, which is also consistent with the corresponding change in the valley Chern number. Moreover, the slope of the SROM demonstrates a sudden jump by one unit of $e/h$ (where $e$ is the electronic charge and $h$ is the Planck's constant) across the TPT, which is also in agreement with the corresponding change in the spin Chern number. It is further seen that a definite spin-valley optical selection rule governs the circular dichroism. The $k$-resolved degree of the optical polarization and the low-frequency differential optical absorbance manifest sign change across the TPT. We discuss experimentally viable signatures of different quantum spin-Hall phases in the optical absorbance.
\end{abstract}
\maketitle

\section{Introduction}
The magnetization of a solid is formally defined as the magnetic dipole moment per unit volume. It is comprised of two parts: the spin magnetization and the orbital magnetization (OM). While the spin magnetization can be determined from the first principles with high accuracy, on the contrary, the precise estimation of the OM has been a persistent issue over the years.
The calculation of the OM in a finite system is straightforward. However, the difficulties in the calculation of the OM for the extended/periodic systems appear owing to the fact that the position operator becomes ill defined in this context.
Earlier attempts were made to develop a theory of the OM in periodic systems based on the linear response methods [\onlinecite{{Morb_theory1},{Morb_theory2},{Morb_theory3},{Morb_theory4},{Morb_theory5}}] which deals only with the change in the magnetization instead of the magnetization itself. Xiao {\it et al.} [\onlinecite{Morb_semi1}] derived a formula for the OM based on Berry phase controlled semiclassical dynamics of Bloch electrons.
Thonhauser {\it et al.} [\onlinecite{Morb_modern}] developed a modern theory of the OM in crystalline insulators using Wannier functions [\onlinecite{{Morb_wan1},{Morb_wan2}}] which mimics an analogous derivation of the electric polarization [\onlinecite{El_Pol1,El_Pol2,El_Pol3}]. This theory was further generalized to the cases of metals and Chern insulators [\onlinecite{{Morb_chern1},{Morb_chern2},{Morb_chern3}}]. Furthermore, Shi {\it et al.} [\onlinecite{Morb_quant}] derived a full quantum mechanical expression of the OM for interacting systems using standard perturbation techniques.

The general formula for the OM adds a crucial Berry phase correction term to the conventional OM. This topological correction term can be regarded as a bulk property of the system. Over the years, the topological characteristics of the OM has been investigated in several systems including Haldane model [\onlinecite{Morb_chern1, disorder_Haldane}], Kagome systems [\onlinecite{Kagome, Kagome_mag}], antiferromagnetic distorted 
face-centered-cubic lattice [\onlinecite{fcc}], electron-hole doped  semiconductors [\onlinecite{Semicon}], transition metal dichalcogenides [\onlinecite{TMDC1, TMDC2}], 
bilayer Chern system [\onlinecite{bi_chern1, bi_chern2}], etc. Recently, it is proposed that the OM can play a fundamental role in a circularly polarized light-controlled topological memory device [\onlinecite{Topo_memory}].

Recent years have witnessed a tremendous interest in the 
$\alpha$-$T_3$ system [\onlinecite{Illes_Thesis}]. With the variation of a parameter $\alpha\in(0,1)$, the $\alpha$-$T_3$ lattice model provides a bridge between the honeycomb structure of 
graphene ($\alpha=0$) and the dice lattice ($\alpha=1$) [\onlinecite{{dice1},{dice2},{dice3}}]. It can be realized experimentally in the heterostructures as well as in the optical lattice avenues as proposed in some earlier studies [\onlinecite{Dice_Real,Dice_Opt,alp_T3_real}]. A nearest-neighbor tight-binding calculation reveals that the $\alpha$-$T_3$ lattice hosts massless quasiparticles which obey the Dirac-Weyl equation  with a generalized 
$\alpha$-dependent pseudospin. The existence of a zero energy flat band and the $\alpha$-dependent Berry phase make the model more interesting which motivated people for extensive investigations in various directions [\onlinecite{Mag_Suscp,Klein1,Klein2,T3_Hall1,T3_Hall2,Weiss,ZB,Plasmon1,Plasmon2,
Plasmon3,Plasmon4,Mag_Opt1,Mag_Opt2,Mag_Opt3,Mag_Opt4,RKKY1,RKKY2,Min_Con, Ghosh_Topo,GVHET3,Imp_T3,StrainT3,StrainT32,GirishS,RingT3}].
The $\alpha$-$T_3$ model exhibits nontrivial topological features in 
\textcolor{blue}{the} presence of a time periodic radiation [\onlinecite{{Flo1},{Flo2},{Flo3},{Flo4},{Flo5}}]. Particularly, it undergoes a topological phase transition(TPT), when exposed to a circularly polarized off-resonant light [\onlinecite{Flo2}], delineated by a change in the Chern number from $C=1$ to $C=2$ across $\alpha=1/\sqrt{2}$. A recent work [\onlinecite{Flo5}] has unveiled the signatures of this TPT in the OM. It is explicitly shown therein that the OM becomes linear when the chemical potential varies in the band gap(s) of the energy spectrum. More specifically, the slope of the OM changes by one unit of $e/h$ across $\alpha=1/\sqrt{2}$.

Recently, Wang and Liu [\onlinecite{Spin_Hall_Phase}] demonstrated that the $\alpha$-$T_3$ lattice model supports quantum spin-Hall phases when the spin-orbit interaction (SOI) of 
Kane-Mele type [\onlinecite{Kane_Mele}] is taken into account. Additionally, it is reported therein that the underlying system undergoes a TPT characterized by a change in the spin Chern number from $C_s=1$ to $C_s=2$ across $\alpha=0.5$. At this juncture, it would be interesting to look at the traces of the TPT in some physical observables which might be detected experimentally. Motivated by this we study the behavior of the Berry curvature, the orbital magnetic moment(OMM), and the OM of a spin-orbit coupled 
$\alpha$-$T_3$ system. 
It was argued [\onlinecite{CD1,CD2,CD3,CD4,CD5}] that the circular dichroism could be considered as a possible probe for the OM. Souza and Vanderbilt [\onlinecite{CD4}] derived an exact sum rule that connects the frequency integral of the imaginary part of the optical conductivity to a gauge invariant part of the total OM. 
Yao {\it et al.} [\onlinecite{CD_Yao}] proposed an optical scheme to measure the OM via dichoric sum rules by establishing a relation between the $k$-resolved degree of optical polarization and orbital magnetic moment for a two-band system.
M. Ezawa [\onlinecite{CD_Ezawa}] suggested that the TPT in silicene can be detected via spin-valley selective circular dichroism. An experimental evidence of valley-selective circular dichroism was confirmed in monolayer molybdenum disulphide by Cao {\it et al.} [\onlinecite{CD_Expt}]. Therefore, the connection between the OM and the circular dichroism could provide an optical way of detecting the TPT.

Our observations are as follows.
Both the Berry curvature and the OMM change their respective signs discontinuously across the TPT. The SOI essentially introduces three well separated forbidden gaps in the energy spectrum when $0<\alpha<1$. The topological features in the OM are uncovered when the chemical potential scans those forbidden gaps. As the total OM, i.e. spin and valley integrated OM, vanishes because of the time-reversal(TR) symmetry, we study both the valley-resolved OM and the spin-resolved OM, separately. Both the valley-resolved OM and the spin-resolved OM vary linearly with the chemical potential in the forbidden gaps. The slope of the valley-resolved OM changes sign abruptly across the TPT mimicking the corresponding behavior of the valley Chern number. On the other hand, the slope of the spin-resolved OM in the gap for 
$\alpha>0.5$ differs from that for 
$\alpha<0.5$ by one unit of $e/h$. This fact is consistent with the corresponding change in the spin Chern number.
Finally, we demonstrate a possible optical way of detecting this TPT via the circular dichroism associated with interband optical absorption. The essential features of the circular dichroism are drastically different across the TPT. Specifically, the $k$-resolved degree of the optical polarization and low-frequency differential optical absorbance change their respective signs across the TPT.

The rest of the paper is organized as follows. In Sec. \ref{Sec2}, we briefly mention the salient characteristics of the energy band structure of the spin-orbit coupled $\alpha$-$T_3$ lattice. We discuss the behavior of the Berry curvature as well as the OMM in Sec. \ref{Sec3}. The topological signatures in the OM are discussed in Sec. \ref{Sec4}. We provide a detailed discussion on the circular dichroism in Sec. \ref{Sec5}. We summarize our results in Sec. \ref{Sec6}.

\section{Spin-orbit coupled $\alpha$-$T_3$ lattice}\label{Sec2}
\begin{figure}[h!]
\centering
\includegraphics[width=7cm, height=6.5cm]{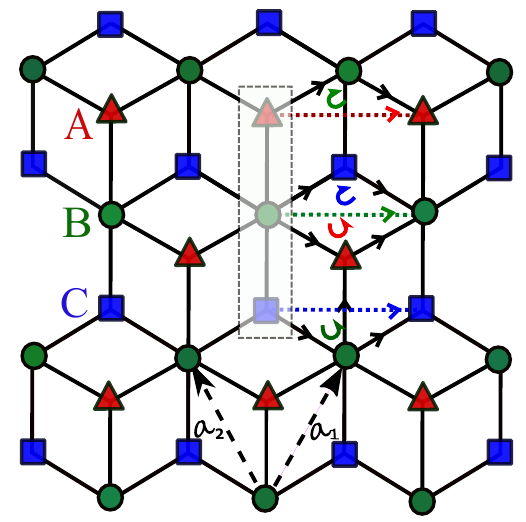}
\caption{The geometry of the
$\alpha$-$T_3$ lattice is shown here. There are three sites, namely, 
$A$, $B$, and $C$ in an unit cell (denoted by the dashed rectangle). The rim sites $A$ and $C$, both have three NN sites while the hub site $B$ has six NNs. The NN connections are denoted by solid lines while the NNN hoping paths are marked by dotted lines. An electron can hop from a given site to its NNN site via two NN paths either in a clockwise or in a counter clockwise sense as viewed from the positive $z$-axis (outward normal to the lattice plane).}
\label{fig:Fig_LatSOI}
\end{figure}

Fig. \ref{fig:Fig_LatSOI} describes schematically the geometrical structure of the $\alpha$-$T_3$ lattice. Here, an unit cell has three sites, namely, $A$, $B$, and $C$. The sites $A$ and $B$ make the honeycomb lattice of graphene with nearest-neighbor(NN) hopping strength $t$. The site $C$ sitting at the center of the hexagon connects three surrounding $B$ sites with hopping strength $\alpha t$, where $\alpha\in(0,1)$. The direct hopping between $A$ and $C$ is forbidden. The lattice translation vectors are $\bm a_1=(\sqrt{3}/2,3/2){\rm a}$ and $\bm a_2=(-\sqrt{3}/2,3/2){\rm a}$, where $\rm a$ is the bond length. Likewise graphene, the SOI of Kane-Mele type can also emerge here by considering next nearest-neighbor(NNN) hoppings [\onlinecite{Spin_Hall_Phase}]. We consider the following possible NNN hopping schemes, namely, (I): $A$-$B$-$A$ and $B$-$A$-$B$ and (II): $B$-$C$-$B$ and $C$-$B$-$C$. If the strength of the SOI for (I) is $\lambda$, then we assume that for (II) it is $\alpha\lambda$. 

\begin{figure}[h!]
\centering
\includegraphics[width=9cm, height=5cm]{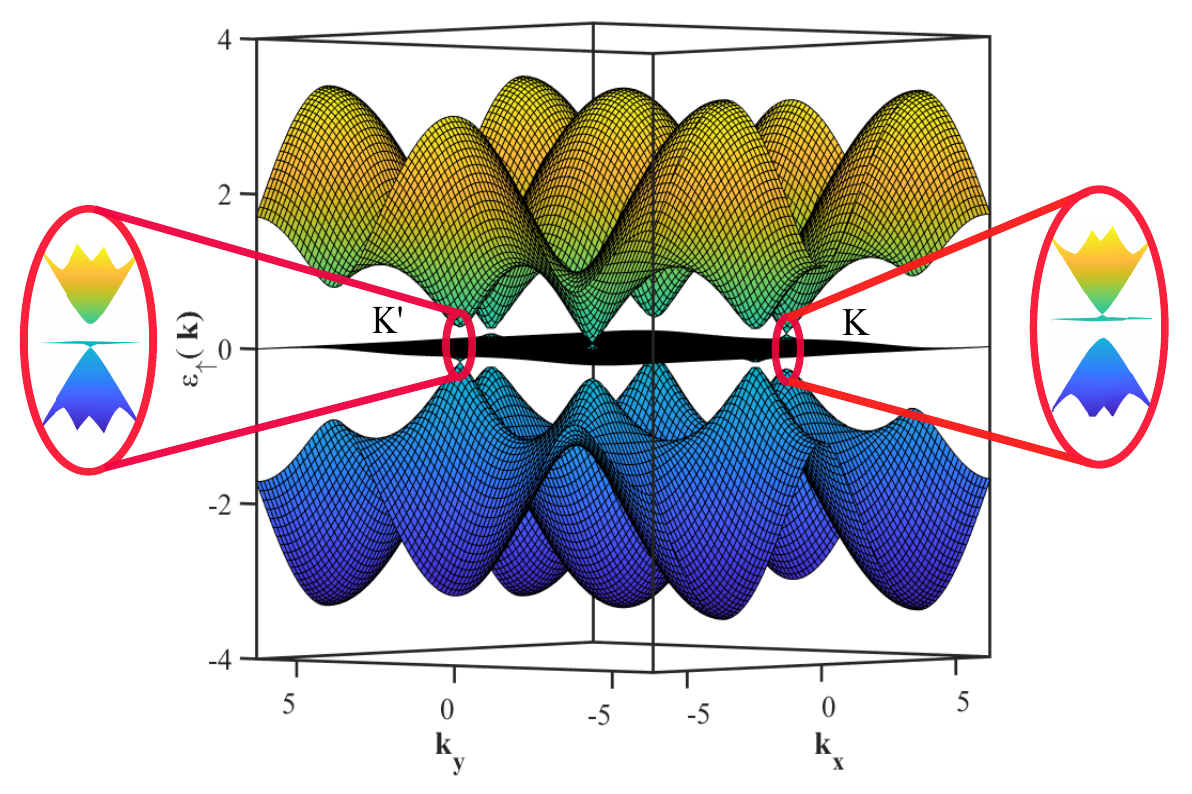}
\caption{The energy dispersion for the spin-up 
($\sigma=\uparrow$) bands over the FBZ considering $\alpha=0.5$. Near the Dirac points i.e. 
$\bm K_\eta=(\eta\frac{4\pi}{3\sqrt{3}\rm a},0)$, the FB is significantly distorted by the SOI. The gap between the CB (VB) and the FB in the $K$ ($K^\prime$) valley closes, as depicted in the zoomed portions.}
\label{fig:Fig_BandS}
\end{figure}

The tight-binding Hamiltonian which incorporates both the NN and the NNN hoppings can be written as
\begin{eqnarray}\label{Ham_tight}
H&=&-t\sum_{\la i,j\ra,\sigma}\mathcal{C}^{\dagger}_{iA,\sigma}\mathcal{C}_{jB,\sigma}-\alpha t\sum_{\la i,j\ra,\sigma}
\mathcal{C}^{\dagger}_{iB,\sigma}\mathcal{C}_{jC,\sigma}\nonumber\\
&+&\frac{i}{3\sqrt{3}}\sum_{\la\la i, j\ra\ra,\gamma, \sigma,\sigma^\prime} \nu_{ij}\lambda_{\gamma} \mathcal{C}^{\dagger}_{i\gamma,\sigma}\sigma_z\mathcal{C}_{j\gamma,\sigma^\prime},
\end{eqnarray}
where $\mathcal{C}^{\dagger}_{i\gamma,\sigma}\,(\mathcal{C}_{i\gamma,\sigma})$  creates (annihilates) an electron with spin orientation 
$\sigma=\uparrow,\, \downarrow$ at the lattice site $i\gamma$ ($\gamma=A, B, C$). The symbol $\la i,j\ra$ ($\la\la i,j\ra\ra$) represents NN (NNN) hopping process occuring between different (same) sublattice.
First two terms of Eq. (\ref{Ham_tight}) are the NN $A$-$B$ and $B$-$C$ hoppings, respectively. The last term describes the NNN hopping schemes  according to (I) and (II) with $(\lambda_A, \lambda_B, \lambda_C)=(-\lambda, \lambda-\alpha\lambda, \lambda)$. Further,
$\nu_{ij}=-1(+1)$ if the NNN hopping is clockwise(counter clockwise) with respect to the positive $z$-axis.
$\sigma_{z}$ is the $z$-component of the usual Pauli spin operator representing electron's real spin. In momentum space, the Hamiltonian in Eq.\,(\ref{Ham_tight}) can be represented in the basis $(\mathcal{C}_{A,\bm k,\uparrow}, \mathcal{C}_{B,\bm k,\uparrow}, \mathcal{C}_{C,\bm k,\uparrow}, \mathcal{C}_{A,\bm k,\downarrow}, \mathcal{C}_{B,\bm k,\downarrow}, \mathcal{C}_{C,\bm k,\downarrow})$ as
\begin{eqnarray}
H_{6\times 6}(\bm k)=
\begin{pmatrix}
\mathcal{H}_\sigma(\bm k) & 0_{3\times 3}\\
0_{3\times 3} & \mathcal{H}_{-\sigma}(\bm k)
\end{pmatrix},
\end{eqnarray}
where $0_{3\times 3}$ is the $3\times3$ null matrix and 
$\mathcal{H}_\sigma(\bm k)$ is given by
\begin{eqnarray}\label{Spin_Ham}
\mathcal{H}_\sigma(\bm{k})&=&
\begin{pmatrix}
0& -t{\rm g}_{\bm{k}} & 0\\
-t{\rm g}_{\bm{k}}^\ast &0 & -\alpha t{\rm g}_{\bm{k}}\\
0& -\alpha t {\rm g}_{\bm{k}}^\ast &0
\end{pmatrix}\nonumber\\
&+&\frac{2\sigma {\rm Im}({\rm g}_{0\bm k})}{3\sqrt{3}}
\begin{pmatrix}
-\lambda & 0 & 0\\
0 &(1-\alpha)\lambda & 0\\
0 & 0 & \alpha\lambda
\end{pmatrix}.
\end{eqnarray}
Here, ${\rm g}_{\bm k}=1+e^{i\bm k\cdot {\bm a}_1}+e^{i\bm k\cdot {\bm a}_2}$ and 
${\rm g}_{0\bm k}=e^{i\bm k\cdot {\bm a}_1}-e^{i\bm k\cdot {\bm a}_2}
-e^{i\bm k\cdot ({\bm a}_1-{\bm a}_2)}$. Note that the SOI cannot mix the spin-up($\sigma=\uparrow$) states with the spin-down($\sigma=\downarrow$) states.  The Hamiltonian $\mathcal{H}_{\sigma}(\bm k)$ for a particular spin orientation $\sigma$, given in 
Eq.\,(\ref{Spin_Ham}), can be diagonalized exactly. The energy dispersion of spin-up 
bands($\sigma=\uparrow$) over the entire first Brillouin zone(FBZ) is shown in 
Fig.\,\ref{fig:Fig_BandS} considering $\alpha=0.5$. The band structure consists of three bands, namely, the conduction band(CB), the flat band(FB), and the valence band(VB). In general, the SOI causes a distortion of the FB near the Dirac points or the valleys, namely, $K$ and $K^\prime$. It also makes the energy spectrum gapped for all values of $\alpha$ except $\alpha=0.5$. Therefore, the degeneracies of the energy bands at the Dirac points, corresponding to the
$\alpha$-$T_3$ lattice without SOI, are being lifted up. Interestingly, the gap between the CB and the FB in the $K$ valley and that between the FB and the VB in the $K^\prime$ valley close when $\alpha$ becomes $\alpha=0.5$. As the system is TR invariant, the role of the spin-up bands in the $K$ valley would be replaced by that of the spin-down bands in the $K^\prime$ valley and vice-versa.   

In the vicinity of a particular Dirac point characterized by the valley index $\eta=\pm 1$, the Hamiltonian in Eq.\,(\ref{Spin_Ham}) is further reduced to 
\begin{eqnarray}\label{Ham_Valley}
H_\sigma^\eta(\bm{k})&=&
\begin{pmatrix}
0&f_{\bm{k}} \cos\phi&0\\
f_{\bm{k}}^* \cos\phi&0&f_{\bm{k}}\sin\phi\\
0&f_{\bm{k}}^*\sin\phi&0
\end{pmatrix}\nonumber\\
&-&{\lambda\eta\sigma}
\begin{pmatrix}
\cos\phi&0&0\\
0&{\sin\phi-\cos\phi}&0\\
0&0&{-\sin\phi}
\end{pmatrix},
\end{eqnarray}
where $\tan\phi=\alpha$ and 
$f_{\bm{k}}=\hbar v_F(\eta k_x + ik_y)$
with $v_F=3{\rm a}t/2$ being the Fermi
velocity. Here, the wave vector $\bm k$ is measured from the Dirac point i.e. 
$\bm k\to \bm k-\bm K_\eta$. Note that $H_\sigma^\eta(\bm{k})$ in 
Eq.\,(\ref{Ham_Valley}) is rescaled by $\cos\phi$ just like the case without the SOI [\onlinecite{Mag_Suscp}].
The energy eigenvalue is obtained as
\begin{eqnarray}\label{Eigen_Energy}
\varepsilon_{\eta,\sigma}^n({\bm k})=2\sqrt{\frac{-p}{3}} \cos \Bigg[\frac{1}{3}\arccos\Bigg(\frac{3q}{2p}\sqrt{\frac{-3}{p}}\Bigg)-\frac{2\pi n}{3}\Bigg],
\end{eqnarray}
where $p=\lambda^2\sin(2\phi)/2-(\lambda^2+\hbar^2v_F^2k^2)$
and $q=\lambda\eta\sigma(\lambda^2+\hbar^2v_F^2k^2)(\cos\phi-\sin\phi)\sin(2\phi)/2$.
Here, $n=0,1,$ and $2$ correspond to the CB, the FB and the VB, respectively.
The normalized eigenspinor is given by
\begin{eqnarray}
\psi_{\eta,\sigma}^n(\bm k)=N_{\eta,\sigma}^n(\bm k)
\begin{pmatrix}
\frac{f_{\bm{k}}\cos\phi}{\varepsilon_{\eta,\sigma}^n+\lambda\eta\sigma\cos\phi}\\
\\
1\\
\\
\frac{f_{\bm{k}}^*\sin\phi}{\varepsilon_{\eta,\sigma}^n-\lambda\eta\sigma\sin\phi}
\end{pmatrix}\frac{e^{i\bm k\cdot \bm r}}{\sqrt{S}},
\end{eqnarray}
where the normalization function is given by
\begin{eqnarray}
N_{\eta,\sigma}^n(\bm k)=\Bigg[1 &+& \Bigg(\frac{\vert f_{\bm{k}}\vert\cos\phi}{\varepsilon_{\eta,\sigma}^n+\lambda\eta\sigma\cos\phi}\Bigg)^2\nonumber\\
&+&\Bigg(\frac{\vert f_{\bm{k}}\vert\sin\phi}{\varepsilon_{\eta,\sigma}^n-\lambda\eta\sigma\sin\phi}\Bigg)^2\Bigg]^{-\frac{1}{2}} :
\end{eqnarray}
and $S$ is the area of the sample.
As mentioned earlier that the energy bands, $\varepsilon_{\eta,\sigma}^n({\bm k})$, respect the TR symmetry irrespective of the value of $\alpha$. However, broken spatial inversion(SI) symmetry makes the bands spin and valley polarized at an intermediate value of $\alpha$, i.e., $0<\alpha<1$. 
The SI symmetry is restored in the limiting cases i.e., $\alpha=0$ and $\alpha=1$. Consequently the energy bands become spin and valley degenerate i.e. $\varepsilon_{\eta,\sigma}^n=\varepsilon_{\tilde{\eta},\tilde{\sigma}}^n$, where $\tilde{\eta}=-\eta$ and $\tilde{\sigma}=-\sigma$.

\section{Berry Curvature and Orbital Magnetic moment}\label{Sec3}
In this section, we systematically discuss the behavior of the Berry curvature and the orbital magnetic moment(OMM) and the topological features therein.

\subsection{Berry curvature}
It is a known fact that only the transverse component of the Berry curvature associated with a two-dimensional system survives.
The $z$-component of the Berry curvature of a given Bloch band, characterized by the indices ($n,\eta,\sigma$), is expressed in the following gauge-invariant form [\onlinecite{Berry_Rev}]
\begin{eqnarray}\label{Berry_C}
{{\Omega}}^{n}_{\eta,\sigma}({\bm k})=-2\hbar^2\, {\rm Im}\sum\limits_{\substack{n^\prime\\
(n\neq n^\prime)}}
{\frac{{\langle u^n_{\eta,\sigma}\vert v_x\vert{u^{n^\prime}_{\eta,\sigma}}\rangle\langle u^{n^\prime}_{\eta,\sigma}\vert v_y \vert{u^n_{\eta,\sigma}}\rangle}}{\left[\varepsilon^{n}_{\eta,\sigma}(\bm k)-\varepsilon^{n^\prime}_{\eta,\sigma}(\bm k)\right]^2}},
\end{eqnarray}
where $u^n_{\eta,\sigma}\equiv \vert u^n_{\eta,\sigma}(\bm k)\ra=\sqrt{S}e^{-i\bm k\cdot \bm r}\vert\psi^n_{\eta,\sigma}(\bm{k})\ra$ and 
$v_i=\hbar^{-1}\nabla_{\bm k_i}H_\sigma^\eta(\bm k)$ is the velocity operator along a particular direction $i=x,y$.

The Berry curvature obeys following symmetry properties: 
$\Omega_{\eta,\sigma}^n(\bm{k})=\Omega_{\tilde{\eta},\sigma}^n(\bm k)$ under the SI and 
$\Omega_{\eta,\sigma}^n(\bm{k})=-\Omega_{\tilde{\eta},\tilde{\sigma}}^n(\bm k)$ under the TR operations. The breaking of any of these symmetries gives rise to a non-vanishing Berry curvature [\onlinecite{ValleyCon_Graph}]. 
Due to the combined effect of the TR and the SI symmetries, the Berry curvature would vanish when $\alpha=0$(graphene) and $\alpha=1$(dice lattice). This can be understood in the following way. We obtain the Berry curvature associated with a given band characterized by the indices $(n,\eta,\sigma)$ analytically as
\begin{eqnarray}\label{Berry_dice}
\Omega_{\eta,\sigma}^n({\bm k})=\frac{(n-1)\sigma\lambda\hbar^2v_F^2}{\sqrt{2}(\hbar^2 v_F^2k^2+\frac{\lambda^2}{2})^{3/2}}.
\end{eqnarray}
and 
\begin{eqnarray}\label{Berry_graphene}
\Omega_{\eta,\sigma}^n({\bm k})=\frac{(n-1)\sigma\lambda\hbar^2v_F^2}{2(\hbar^2 v_F^2k^2+\lambda^2)^{3/2}}.
\end{eqnarray}
for $\alpha=1$ and $\alpha=0$, respectively. Note that $n=1$ in Eq. (\ref{Berry_graphene}) is excluded because graphene does not possess any FB. The CB and the VB are denoted by $n=0$, and $n=2$, respectively. As evident from Eq. (\ref{Berry_dice}) and Eq. (\ref{Berry_graphene}), the Berry curvature does not depend on the valley index $\eta$ in both cases $\alpha=0$ and $\alpha=1$ because of the SI symmetry. As the energy band at a given valley is spin degenerate in each cases, a summation over $\sigma$ would give rise to vanishing Berry curvature.

\begin{figure}[h!]
\centering
 \includegraphics[width=10.5cm, height=6.5cm]{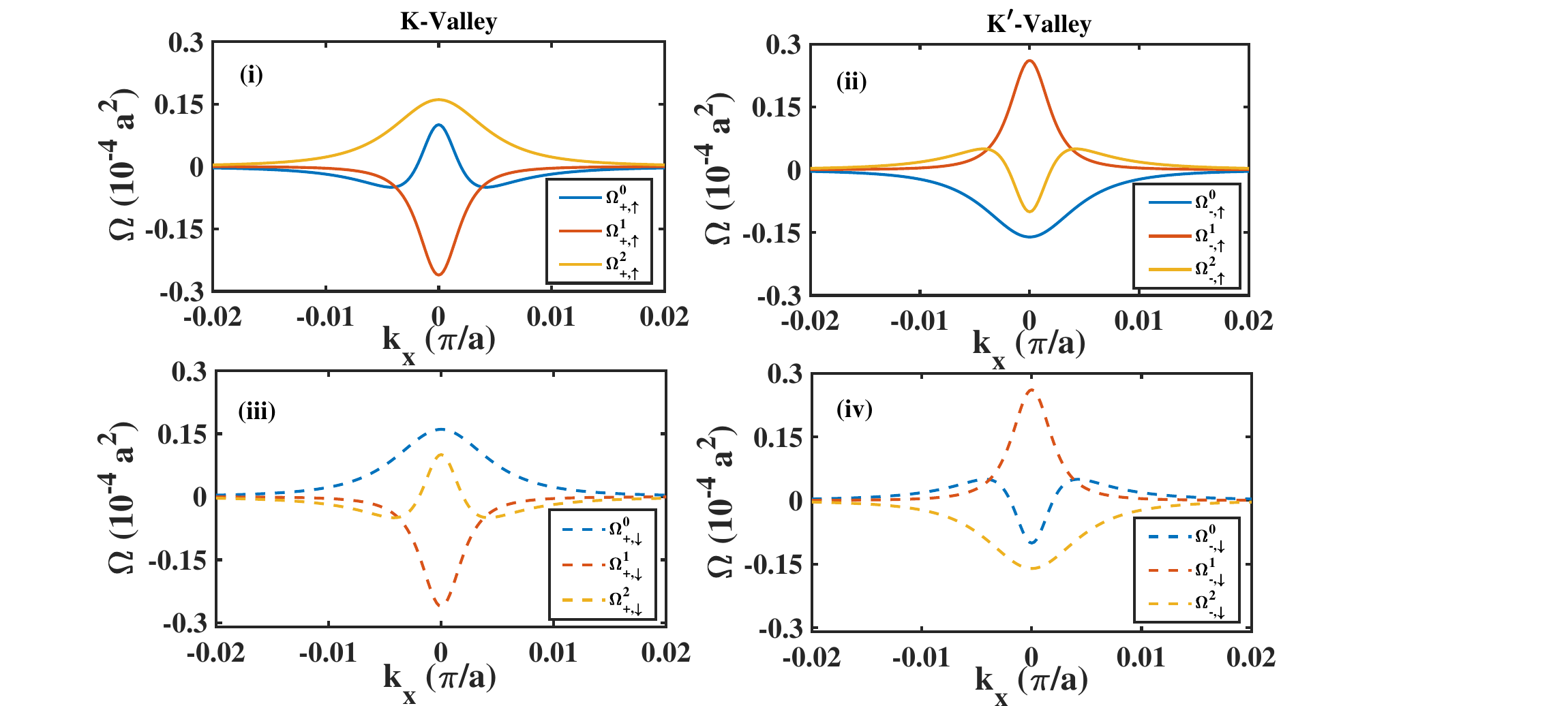}
 \caption{The momentum space distributions of the Berry curvature are shown for the $K$ and the $K^\prime$ valleys. Here, we consider $\alpha=0.3$ and 
$\lambda=50$ meV. The Berry curvature associated with different energy bands, except the FB, are spin-polarized. In addition, the Berry curvature of a given band is valley-polarized.}
\label{fig:Fig_BerryC}
\end{figure}

Fig. \ref{fig:Fig_BerryC} depicts the behavior of the Berry curvature for both spin orientations in the $K$ and the $K^{\prime}$ valleys considering 
$\alpha=0.3$. The Berry curvature associated with different bands are mostly concentrated around the Dirac points. The broken SI symmetry has the following consequences. First of all it permits the FB to acquire some finite Berry curvature. A careful inspection on Fig. \ref{fig:Fig_BerryC} reveals that
$\Omega_{\eta,\uparrow}^{0(2)}(\bm k)=\Omega_{\eta,\downarrow}^{2(0)}(\bm k)$ and $\Omega_{\eta,\uparrow}^1(\bm k)=\Omega_{\eta,\downarrow}^1(\bm k)$. Therefore, the Berry curvature associated with the FB ($n=1$) at a particular valley is spin-degenerate. However, the Berry curvatures associated with the CB($n=0$) and the VB ($n=2$) are spin-polarized.
One can also have following observations from
Fig. \ref{fig:Fig_BerryC} : $\Omega_{+,\sigma}^{0(2)}(\bm k)=-\Omega_{-,\sigma}^{2(0)}(\bm k)$ and $\Omega_{+,\sigma}^1(\bm k)=-\Omega_{-,\sigma}^1(\bm k)$. Therefore, the Berry curvature associated with a particular band having a definite spin orientation is valley-polarized. It is also clear from Fig. \ref{fig:Fig_BerryC} that the Berry curvature associated with a particular band $n$ has opposite signs in the two valleys for opposite spin orientations that is, ${\Omega}_{\eta,\sigma}^n(\bm{k})=-{\Omega}_{\tilde{\eta},\tilde{\sigma}}^n(\bm{k})$, as required by the TR symmetry. The total Berry curvature, i.e. sum of the individual contributions due to different energy bands characterized by both the spin and valley indices, vanishes, which is a requirement for the local conservation of the Berry curvature.

\begin{figure}[h!]
\centering
\includegraphics[width=8.5cm, height=6.5cm]{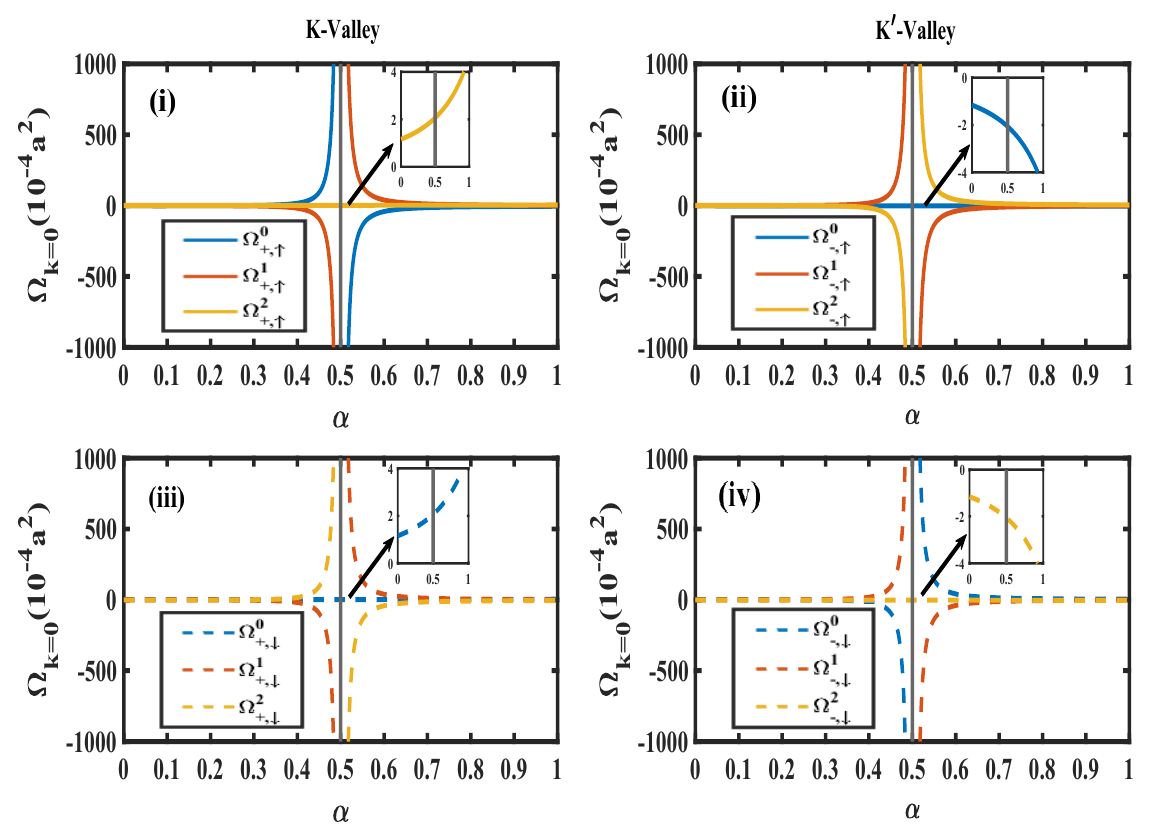}
\caption{The behavior of 
$\Omega(\bm k=0)$ is shown over the full range of $\alpha$. The Berry curvatures associated with the bands which touch each other at the Dirac point  diverge when $\alpha=0.5$. They change their respective signs across 
$\alpha=0.5$, indicating the TPT.}
\label{fig:Fig_BerryCPeak}
\end{figure}

As the Berry curvature has an extremum(peak or dip) around $\bm k=0$, we show the variation of $\Omega(\bm k=0)$ with $\alpha$ in 
Fig. \ref{fig:Fig_BerryCPeak}. In the $K$ valley[Figs. \ref{fig:Fig_BerryCPeak}(i) and \ref{fig:Fig_BerryCPeak}(iii)], $\Omega_{+,\uparrow}^2$ and $\Omega_{+,\downarrow}^0$ increase monotonically with $\alpha$. However, $\Omega_{+,\uparrow}^0$, $\Omega_{+,\uparrow}^1$, 
$\Omega_{+,\downarrow}^1$, and $\Omega_{+,\downarrow}^2$ diverge when $\alpha=0.5$. The gap closing between the energy bands $\varepsilon_{+,\uparrow}^0$ and  $\varepsilon_{+,\uparrow}^1$ at 
$\alpha=0.5$ causes the divergence of $\Omega_{+,\uparrow}^0$ and $\Omega_{+,\uparrow}^1$. Similarly, the same between $\varepsilon_{+,\downarrow}^1$ and $\varepsilon_{+,\downarrow}^2$ makes $\Omega_{+,\downarrow}^1$, and $\Omega_{+,\downarrow}^2$ to diverge at 
$\alpha=0.5$. Most importantly, we note that $\Omega_{+,\uparrow}^0$, $\Omega_{+,\uparrow}^1$, 
$\Omega_{+,\downarrow}^1$, and $\Omega_{+,\downarrow}^2$ change their respective signs across $\alpha=0.5$ which can be viewed as a possible indication of the TPT. On the other hand, the $K^\prime$ valley[Figs. \ref{fig:Fig_BerryCPeak}(ii) and \ref{fig:Fig_BerryCPeak}(iv)] hosts analogous physics as per the requirement of the TR symmetry.

\subsection{Orbital magnetic moment}
Usually, a Bloch electron is represented by a wave packet in the semiclassical scenario.
The self rotation of the wave packet about the center of mass generates an intrinsic OMM whose general expression is given by [\onlinecite{O_Mag_Moment}]
\begin{eqnarray}
\bm{m}_{\eta,\sigma}^n(\bm{k})=-\frac{ie}{2\hbar}\langle\nabla_{\bm{k}} 
u_{\eta,\sigma}^n\vert\times[H_\sigma^\eta(\bm{k})-\varepsilon_{\eta,\sigma}^n(\bm{k})]\vert\nabla_{\bm{k}}u_{\eta,\sigma}^n\rangle.
\end{eqnarray} 

The $z$-component of the OMM can be written as
\begin{eqnarray}
{m}_{\eta,\sigma}^n(\bm{k})=-\hbar e\sum\limits_{\substack{n^\prime\\
(n\neq n^\prime)}}
{\frac{{\langle u_{\eta,\sigma}^n\vert v_x\vert{u^{n^\prime}_{\eta,\sigma}}\rangle\langle u^{n^\prime}_{\eta,\sigma}\vert v_y \vert{u^n_{\eta,\sigma}}\rangle}}{\left[\varepsilon^n_{\eta,\sigma}(\bm{k})-\varepsilon^{n^\prime}_{\eta,\sigma}(\bm{k})\right]}}.
\end{eqnarray}

The OMM behaves as $m_{\eta,\sigma}^n(\bm{k})=m_{\tilde{\eta},\sigma}^n(\bm k)$ and   
$m_{\eta,\sigma}^n(\bm{k})=-m_{\tilde{\eta},\tilde{\sigma}}^n(\bm k)$   under the SI and the TR operations, respectively. Therefore, the OMM vanishes when the SI and the TR symmetries exist simultaneously. This is exactly the situation corresponding to $\alpha=0$ and $\alpha=1$, as explained in the following. The OMM associated with a band having $(n,\eta, \sigma)$ can be obtained analytically as
\begin{eqnarray}\label{OMM_dice}
m_{\eta,\sigma}^n(\bm{k})=-\frac{\sigma\hbar ev_F^2\lambda}{2\sqrt{2}(\hbar^2 v_F^2k^2+\frac{\lambda^2}{2})}(\delta_{n,0}+2\delta_{n,1}+\delta_{n,2})
\end{eqnarray}
for $\alpha=1$
and 
\begin{eqnarray}\label{OMM_Gr}
m_{\eta,\sigma}^n(\bm{k})=-\frac{\sigma\hbar ev_F^2\lambda}{2(\hbar^2 v_F^2k^2+\lambda^2)}(\delta_{n,0}+\delta_{n,2})
\end{eqnarray}
for $\alpha=0$.
The OMMs in Eq.\,(\ref{OMM_dice}) and Eq.\,(\ref{OMM_Gr}) are valley independent as a result of the SI symmetry. The total OMM of a given band in a particular valley, is obtained by summing over degenerate spin states, that would vanish.

\begin{figure}[h!]
\centering
\includegraphics[width=10.5cm, height=6.5cm]{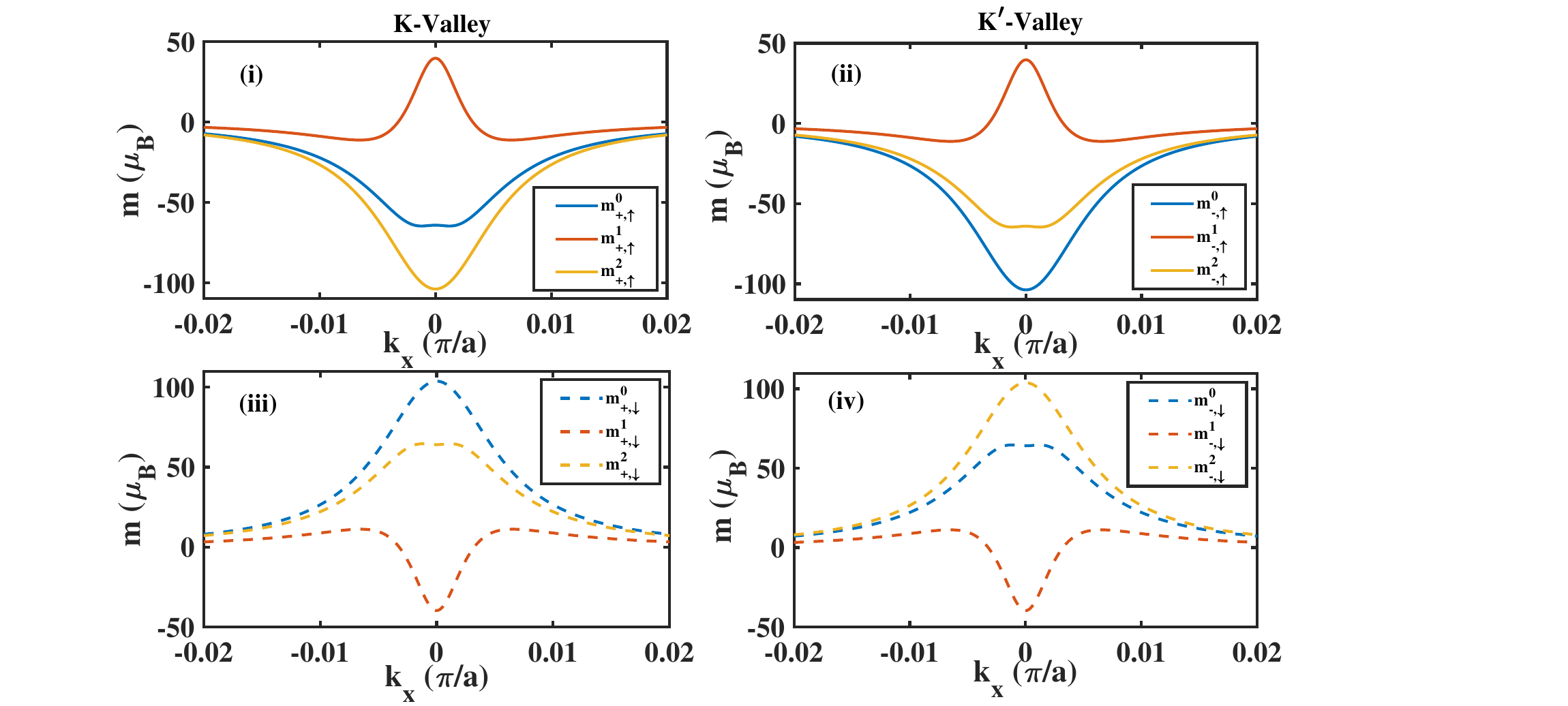}
\caption{The momentum space distributions of the OMM are shown for the $K$ and the $K^\prime$ valleys. Here, we consider 
$\alpha=0.3$ and $\lambda=50$ meV. The OMM associated with a certain energy band in a definite valley is spin-polarized. The OMMs except for the FB are valley-polarized.}
\label{fig:Fig_Magmom}
\end{figure}

The $\bm k$-space distributions of the OMM for both spin orientations around the $K$ and the $K^{\prime}$ valleys are depicted in Fig.\,\ref{fig:Fig_Magmom} considering $\alpha=0.3$. The OMM exhibits an extremum
at $\bm k=0$ likewise $\Omega(\bm k)$. We point out following observations on the behavior of the OMM.
For example, we obtain $m_{\eta,\uparrow}^{0(2)}=-m_{\eta,\downarrow}^{2(0)}$ and $m_{\eta,\uparrow}^{1}=-m_{\eta,\downarrow}^{1}$ which implies that the OMM associated with a given energy band in a particular valley is spin-polarized.
On the other hand, we also have $m_{+,\sigma}^{0(2)}=m_{-,\sigma}^{2(0)}$ and $m_{+,\sigma}^{1}=m_{-,\sigma}^{1}$. The OMM associated with the FB having a particular spin orientation is valley-degenerate. These features appear due to the broken SI symmetry.  Note also that Fig. \ref{fig:Fig_Magmom} (i) and Fig. \ref{fig:Fig_Magmom} (iv) are the TR counterparts because $m_{+,\uparrow}^n=-m_{-,\downarrow}^n$. Same is true for Fig. \ref{fig:Fig_Magmom} (ii) and Fig. \ref{fig:Fig_Magmom} (iii).

\begin{figure}[h!]
\centering
\includegraphics[width=8.5cm, height=6.5cm]{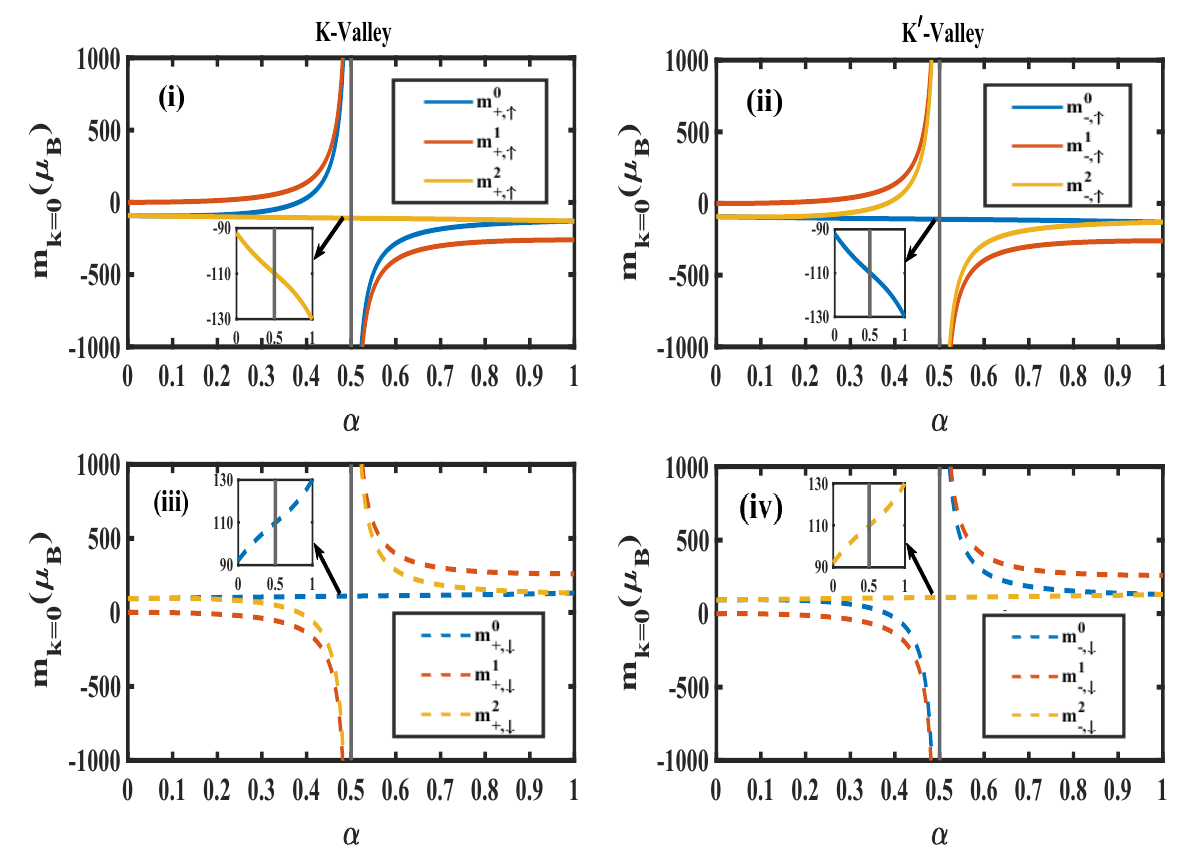}
\caption{The behavior of $m(\bm k=0)$ is shown over the full range of 
$\alpha$. The OMMs associated with the bands which touch each other at the Dirac point diverge when $\alpha=0.5$. They change their respective signs across $\alpha=0.5$, which may be treated as a possible indication of the TPT.}
\label{fig:Fig_Magmom_Peak}
\end{figure}

The behavior of $m(\bm k=0)$ is shown in Fig. \ref{fig:Fig_Magmom_Peak} over the full range of $\alpha$. In the $K$ valley, $m_{+,\uparrow}^2$($m_{+,\downarrow}^0$) decreases(increases) monotonically with $\alpha$. On the other hand, $m_{-,\uparrow}^0$ decreases while $m_{-,\downarrow}^2$ increases with 
$\alpha$ in the $K^\prime$ valley. Interestingly, we note that
$m_{+,\uparrow}^0(m_{-,\uparrow}^2)$ and $m_{+,\uparrow}^1(m_{-,\uparrow}^1)$ change sign from ``+" to ``-" while  $m_{+,\downarrow}^1(m_{-,\downarrow}^1)$ and $m_{+,\downarrow}^2(m_{-,\downarrow}^0)$ change sign from ``-" to ``+" abruptly across $\alpha=0.5$ in the $K(K^\prime)$ valley.  Likewise the Berry curvature, the sign change of the OMM across $\alpha=0.5$ can be considered as a possible signal of a TPT.
 
\section{Orbital Magnetization}\label{Sec4}
In the semiclassical treatment, Xiao $et$ $al.$ [\onlinecite{Morb_semi1}] explicitly showed that the Berry curvature modifies the density of states of the phase space in the presence of a weak external magnetic field $\bm B$ as
\begin{eqnarray}
{\mathcal D}_{\eta,\sigma}^n(\bm k)=\frac{1}{(2\pi)^2}\Big[1+\frac{e}{\hbar}\bm B\cdot{\bs \Omega}_{\eta,\sigma}^n(\bm k)\Big].
\end{eqnarray}
This modified density of states has profound effect in the transport properties [\onlinecite{Berry_Rev, Morb_semi2}]. 
The single particle free energy is given by
\begin{eqnarray}\label{free}
F_{\eta,\sigma}=-\frac{1}{\beta{(2\pi)^2}} \sum_{n}\int d^2{k}\, {\mathcal D}_{\eta,\sigma}^n({\bm k}){\rm ln}\Big[1+e^{-{\beta(E_{\eta,\sigma}^n-\mu)}}\Big].
\end{eqnarray}
Here, $\beta=1/(k_BT)$ with $k_B$ is the Boltzmann constant and $T$ is the temperature,  $E_{\eta,\sigma}^n(\bm{k})=\varepsilon_{\eta,\sigma}^n-{\bm{m}}_{\eta,\sigma}^n\cdot \bm{B}$ is the modified band energy due to the coupling of ${\bm{m}}_{\eta,\sigma}^n$ with $\bm B$, and $\mu$ is the chemical potential.
The OM in a particular valley for a definite spin orientation is given by 
\begin{eqnarray} 
M_{\eta,\sigma}^{\rm orb}=-\frac{\partial{F_{\eta,\sigma}}}{\partial{B}}\Big\vert_{\mu,T},
\end{eqnarray}
which can be further obtained as
\begin{eqnarray}\label{Tot_OM}
M_{\eta,\sigma}^{\rm orb}=M_{\eta,\sigma}^{\rm avg}+M_{\eta,\sigma}^{\rm com}
\end{eqnarray}
where
\begin{eqnarray}\label{M_avg}
M_{\eta,\sigma}^{\rm avg}=\frac{1}{(2\pi)^2}\sum_n\int m_{\eta,\sigma}^n(\bm{k})f_{\eta,\sigma}^n(\bm{k})\,d^2{k}
\end{eqnarray}
and
\begin{eqnarray}\label{M_com}
M_{\eta,\sigma}^{\rm com}=-\frac{e}{2\pi\beta{h}}\sum_n\int \Omega_{\eta,\sigma}^n(\bm{k}) \ln\Big[1-f_{\eta,\sigma}^n(\bm k)\Big]\,d^2{k}.
\end{eqnarray}
Here, $f_{\eta,\sigma}^n(\bm{k})=[\exp\{\beta(\varepsilon_{\eta,\sigma}^n(\bm k)-\mu)\}+1]^{-1}$ is the Fermi-Dirac distribution function. The limits of the integrations in 
Eqs.\,(\ref{M_avg}) and (\ref{M_com}) are such that all the occupied states are only considered. 
The conventional term $M_{\eta,\sigma}^{\rm avg}$  [Eq.\,(\ref{M_avg})] is the thermodynamic average of the OMMs of the carriers. The Berry phase mediated correction term
$M_{\eta,\sigma}^{\rm com}$  [Eq.\,(\ref{M_com})] is topological in nature and associated with the center of mass motion of the wave packet.

\subsection{Valley-resolved Orbital Magnetization}
\begin{figure}[h!]
\centering
\includegraphics[width=8.5cm, height=5cm]{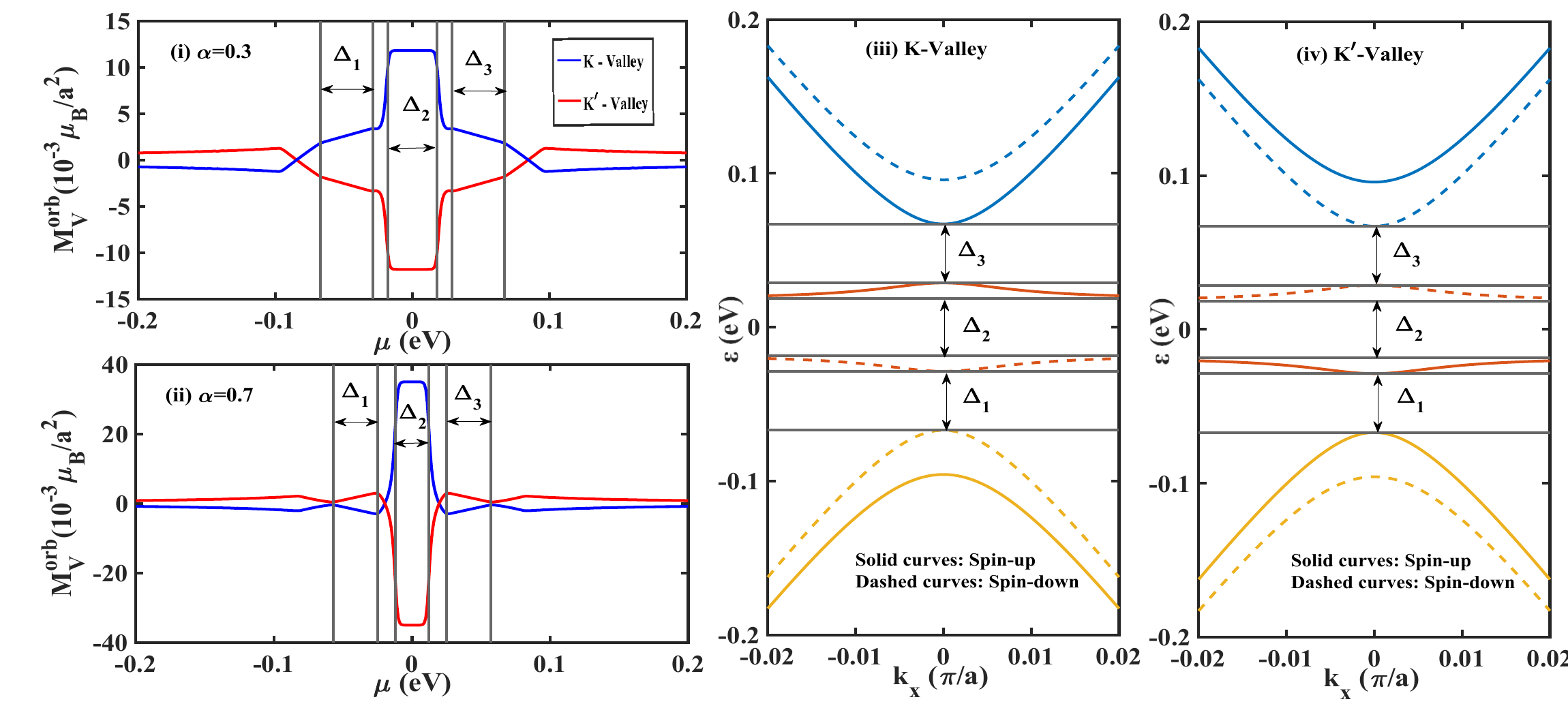}
\caption{The variation of the VOM ($M_{\rm V}^{\rm orb}$) with the chemical potential ($\mu$) for (i) $\alpha=0.3$ and (ii) $\alpha=0.7$. Here, we consider $T=10$ K and $\lambda=100$ meV. 
The spin-polarized energy bands in the $K$ and 
the $K^\prime$ valleys are shown in (iii) and (iv), respectively. The SOI introduces three well separated $\alpha$-dependent windows of forbidden energy, namely $\Delta_1$, $\Delta_2$, and $\Delta_3$ such that $\Delta_1=\Delta_3\neq\Delta_2$. 
For $\alpha=0.3$, we find 
$\Delta_1=0.0383$ eV and $\Delta_2=0.0369$ eV. 
However, we have $\Delta_1=0.0328$ eV and $\Delta_2=0.0234$ eV when 
$\alpha=0.7$.} 
\label{fig:Fig_VOM}
\end{figure}

Let us first define the valley orbital magnetization(VOM), $M_{\rm V}^{\rm orb}$, in a particular valley by adding two spin contributions therein. Therefore, $M_{\rm V}^{\rm orb}$ should mean as $M_{\rm V}^{\rm orb}=M_{+,\uparrow}^{\rm orb}+
M_{+,\downarrow}^{\rm orb}$ for the $K$ valley and $M_{\rm V}^{\rm orb}=M_{-,\uparrow}^{\rm orb}+
M_{-,\downarrow}^{\rm orb}$ for the $K^\prime$ valley.
We sketch the behavior of the VOM with the chemical potential($\mu$)
in Figs. \ref{fig:Fig_VOM}(i) and \ref{fig:Fig_VOM}(ii) for 
$\alpha=0.3$ and $\alpha=0.7$, respectively. The VOM in the
$K$ valley is exactly equal and opposite to that in the 
$K^\prime$ valley as a consequence of the TR symmetry. Therefore, the total OM, i.e., both spin and valley integrated OM, vanishes. 

However, the VOM exhibits some important features that can be connected to the topological properties of the system. Particularly, the behavior of $M_{\rm V}^{\rm orb}$ in the three "windows" of $\mu$, namely 
$\Delta_1$, $\Delta_2$, and $\Delta_3$ are interesting. These windows are identified as the SOI induced forbidden gaps in the energy 
spectrum [Figs. \ref{fig:Fig_VOM}(iii) and \ref{fig:Fig_VOM}(iv)] and their proper meaning are given in Table \ref{Table_Gap}.  
\begin{table}[ht]
\centering
\setlength{\tabcolsep}{12pt}
\renewcommand{\arraystretch}{1.5}
\begin{tabular}{|c|c|c|}\hline

\multicolumn{1}{|c|}{Window} &
\multicolumn{1}{c|}{$K$ valley} &
\multicolumn{1}{c|}{$K^\prime$ valley} \\ \hline
\ \ \ $\Delta_1$ & $\varepsilon_{+,\downarrow}^1-\varepsilon_{+,\downarrow}^2$ & $\varepsilon_{-,\uparrow}^1-\varepsilon_{-,\uparrow}^2$\\
\hline
\ \ \ $\Delta_2$ & $\varepsilon_{+,\uparrow}^1-\varepsilon_{+,\downarrow}^1$ & $\varepsilon_{-,\downarrow}^1-\varepsilon_{-,\uparrow}^1$\\
\hline
\ \ \ $\Delta_3$ & $\varepsilon_{+,\uparrow}^0-\varepsilon_{+,\uparrow}^1$ & $\varepsilon_{-,\downarrow}^0-\varepsilon_{-,\downarrow}^1$\\
\hline
\end{tabular}
\caption{Meaning of the forbidden gaps, $\Delta_1$, $\Delta_2$, and 
$\Delta_3$, in the energy spectrum as depicted in Fig. \ref{fig:Fig_VOM}(iii) and Fig. \ref{fig:Fig_VOM}(iv). Here, the energy gaps $\Delta_1$ and $\Delta_3$ are defined at the Dirac point whereas the gap $\Delta_2$ is defined away from the Dirac point.}
\label{Table_Gap}
\end{table}
Let us now discuss the features of the VOM in the $K$ valley. 
For $\alpha=0.3$, $M_{\rm V}^{\rm orb}$ increases linearly with the chemical potential $\mu$ in the window $\Delta_1$, attains a plateau in $\Delta_2$, and decreases linearly in 
$\Delta_3$. The slope of the linear region in 
$\Delta_1$ is exactly opposite to that in $\Delta_3$. The linear variation of the VOM, when $\mu$ falls in a forbidden gap, is topological in nature which can be explained using the relation[\onlinecite{Morb_chern1}]:
\begin{eqnarray}\label{M_gap}
\frac{dM_{\eta,\sigma}^{\rm orb}}{d\mu}=\frac{e}{h}C_{\eta,\sigma}^{(n)},
\end{eqnarray}
where $C_{\eta,\sigma}^{(n)}$ is the Chern number associated with a definite energy band characterized by the indices $(n,\eta,\sigma)$. It is further defined as
\begin{eqnarray}\label{Eq_Chern}
C_{\eta,\sigma}^{(n)}=\frac{1}{2\pi}\int_{\rm FBZ} d^2k\, 
\Omega_{\eta,\sigma}^n(\bm k).
\end{eqnarray}

In the case of the VOM in the $K$ valley, Eq.\,(\ref{M_gap}) can be modified as 
\begin{eqnarray}\label{M_gap1}
\frac{dM_{\rm V}^{\rm orb}}{d\mu}\Big\vert_{K}=\frac{e}{h}\sum_{n,\sigma}^{\rm occ}C_{+,\sigma}^{(n)}.
\end{eqnarray}
Note that the summation in Eq.\,(\ref{M_gap1}) extends over all the occupied energy bands.
It is clear from Fig. \ref{fig:Fig_VOM}(iii) that spin-up and spin-down VBs i.e. $\varepsilon_{+,\uparrow}^2$ and 
$\varepsilon_{+,\downarrow}^2$ are only occupied when $\mu$ is varied in the window $\Delta_1$. Therefore, Eq.\,(\ref{M_gap1}) dictates that the slope of $M_{\rm V}^{\rm orb}$ is proportional to $C_{+,\uparrow}^{(2)}+C_{+,\downarrow}^{(2)}$. We calculate the slope of VOM from 
Fig. \ref{fig:Fig_VOM}(i) as $0.164$ in units of $e/h$. An explicit calculation of $C_{\eta, \sigma}^{(n)}$ reveals that $C_{+,\uparrow}^{(2)}=0.581$ and $C_{+,\downarrow}^{(2)}=-0.416$, (see Table \ref{Tab_Chern}, Appendix \ref{AppnA}), so that 
$C_{+,\uparrow}^{(2)}+C_{+,\downarrow}^{(2)}=0.165$ which is in excellent agreement with that calculated from Fig. \ref{fig:Fig_VOM}(i). 
When $\mu$ is tuned in the window $\Delta_2$, the VOM becomes flat. 
In this case, three bands, namely, $\varepsilon_{+,\uparrow}^2$, 
$\varepsilon_{+,\downarrow}^2$, and $\varepsilon_{+,\downarrow}^1$ are occupied. The Chern number of the spin-down FB ($\varepsilon_{+,\downarrow}^1$) must be compensated to that of the spin-up and the spin-down VBs together as per the requirement of 
Eq.\,(\ref{M_gap1}) i.e. $C_{+,\downarrow}^{(1)}=-(C_{+,\uparrow}^{(2)}+C_{+,\downarrow}^{(2)})$. This relation is also verified from the direct calculation  in which we obtain $C_{+,\uparrow}^{(1)}=C_{+,\downarrow}^{(1)}=-0.165$ (see Table \ref{Tab_Chern}, Appendix \ref{AppnA}). 
As $\mu$ lies in the window $\Delta_3$, the VOM decreases linearly with a slope of $-0.164$ in units of $e/h$. This slope must be proportional to $C_{+,\uparrow}^{(1)}$ because four bands, namely, 
$\varepsilon_{+,\uparrow}^2$, $\varepsilon_{+,\downarrow}^2$, $\varepsilon_{+,\downarrow}^1$, and $\varepsilon_{+,\uparrow}^1$ are occupied in this case.
On the other hand, when $\alpha=0.7$, the VOM decreases linearly in 
$\Delta_1$, becomes constant in $\Delta_2$, and increases linearly in $\Delta_3$. We find that the slope of the VOM in $\Delta_1$ [calculated from 
Fig.\,\ref{fig:Fig_VOM}(ii)] is $-0.340$ (in units of $e/h$). Here, we calculate Chern numbers explicitly as $C_{+,\uparrow}^{(2)}=0.827$, 
$C_{+,\downarrow}^{(2)}=-1.169$ and $C_{+,\uparrow}^{(1)}=C_{+,\downarrow}^{(1)}=0.342$. The explanation would remain same here as discussed in the case of $\alpha=0.3$. The $K^\prime$ valley also hosts similar physics. 

It is worthy to mention that the linear dependence of $M_V^{\rm orb}$ on $\mu$ in $\Delta_1$ and $\Delta_3$ is entirely due to the Berry phase correction term given in Eq.\,\eqref{M_com}. We have checked (not shown here) the conventional term of the OM [Eq.\,\eqref{M_avg}] attains a plateau when $\mu$ falls in the band gap. The height of the plateau in $\Delta_2$ is larger than that either in $\Delta_1$ or $\Delta_3$.

\begin{figure}[h!]
\centering
\includegraphics[width=8cm, height=6cm]{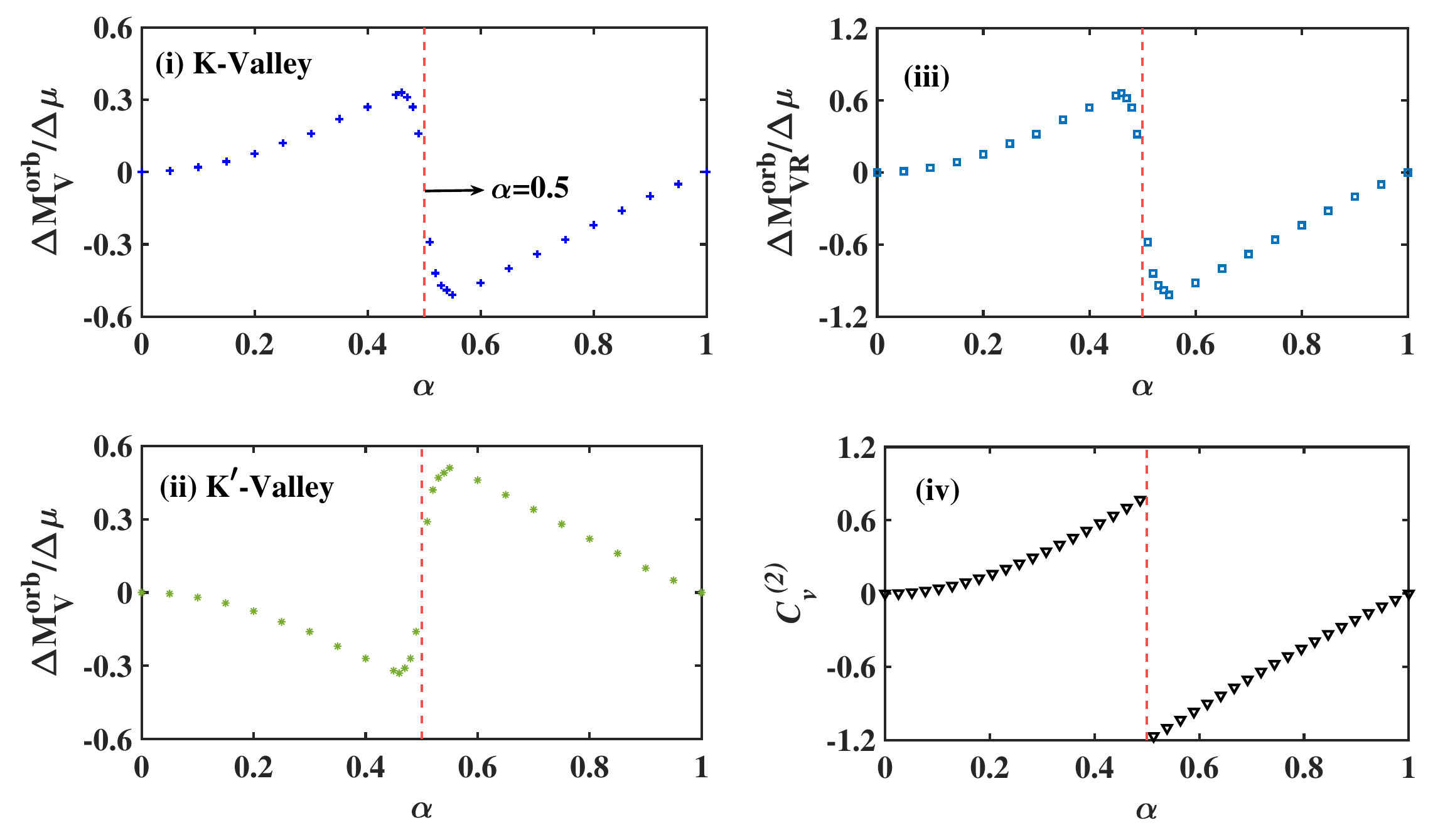}
\caption{(i) The slope of the VOM(in units of $e/h$) in the $K$ valley, (ii) that in the $K^\prime$ valley, and (iii) the slope of the VROM when the chemical potential is tuned in the window $\Delta_1$, for the entire range of $\alpha$. The slope in each case changes sign discontinuously across 
$\alpha=0.5$ to indicate the TPT. (iv) The valley Chern number of the VB, $C_v^{(2)}$, varies in a similar fashion as the slope of the VROM in $\Delta_1$.} 
\label{fig:Fig_VOM_Slope}
\end{figure}

In Fig. \ref{fig:Fig_VOM_Slope}(i) and Fig. \ref{fig:Fig_VOM_Slope}(ii), we depict the nature of the slope of $M_{\rm V}^{\rm orb}$ in the window 
$\Delta_1$ over the full range of $\alpha$ for the $K$ valley and the 
$K^\prime$ valley, respectively. We find that the slope of the VOM in the $K$ valley is exactly opposite to that in the 
$K^\prime$ valley. The slope changes its sign discontinuously across 
$\alpha=0.5$. 

It would be more interesting to see the behavior of the 
valley-resolved orbital magnetization(VROM), defined as $M_{\rm VR}^{\rm orb}=M_{\rm V}^{\rm orb}\vert_K-M_{\rm V}^{\rm orb}\vert_{K^\prime}$, in order to capture the signature of the TPT across $\alpha=0.5$. When $\mu$ lies in
$\Delta_1$, we explicitly find that
\begin{eqnarray}
M_{\rm VR}^{\rm orb}=\frac{e}{h}C_v^{(2)},
\end{eqnarray} 
where $C_v^{(2)}$ is the valley Chern number associated with the VB, defined as 
\begin{eqnarray}
C_v^{(2)}=C_{+,\uparrow}^{(2)}+C_{+,\downarrow}^{(2)}-C_{-,\uparrow}^{(2)}-
C_{-,\downarrow}^{(2)}.
\end{eqnarray}
In Fig. \ref{fig:Fig_VOM_Slope}(iii), the slope of the VROM versus the chemical potential in $\Delta_1$ is depicted over the entire range of $\alpha$. The slope changes its sign abruptly across $\alpha=0.5$ to indicate the TPT. This behavior of $M_{\rm VR}^{\rm orb}$ is consistent with the corresponding nature of the valley Chern number as portrayed in Fig.\,\ref{fig:Fig_VOM_Slope}(iv).  

\subsection{Spin-resolved Orbital Magnetization}
For a better visualization of the TPT across $\alpha=0.5$, we now study the 
$\mu$ dependence of the spin-resolved orbital magnetization (SROM) defined as
\begin{eqnarray}
M_{\rm SR}^{\rm orb}=\frac{1}{2}\Big(M_{\rm +,\uparrow}^{\rm orb}-M_{\rm +,\downarrow}^{\rm orb}+M_{\rm -,\uparrow}^{\rm orb}-
M_{\rm -,\downarrow}^{\rm orb}\Big).
\end{eqnarray}

\begin{figure}[h!]
\centering
\includegraphics[width=8.5cm, height=5cm]{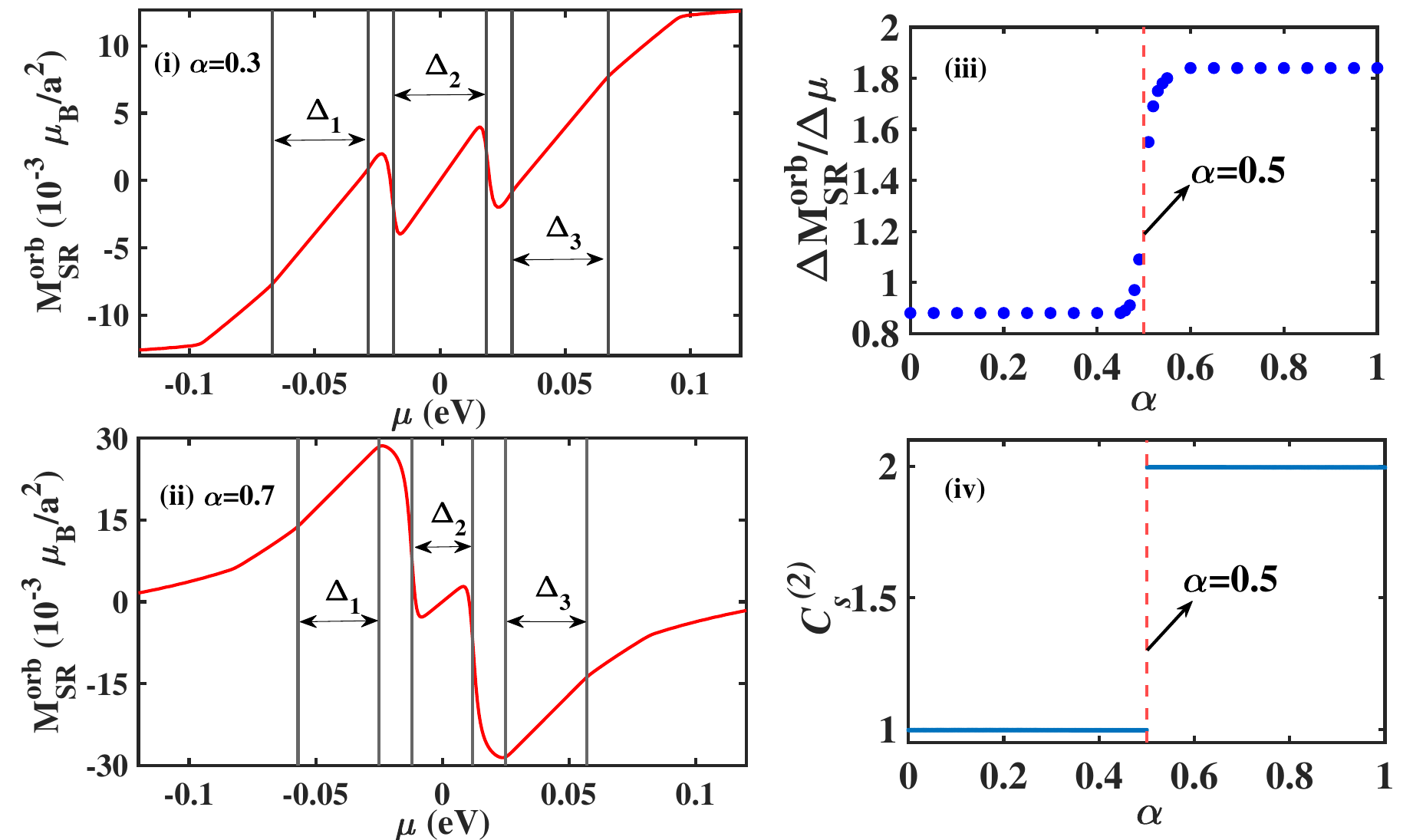}
\caption{The dependence of the SROM on $\mu$ is shown for (i) $\alpha=0.3$ and (ii) $\alpha=0.7$. (iii) The slope of the SROM (in units of $e/h$) in the window $\Delta_1$ over the entire range of 
$\alpha$. The slope changes discontinuously across $\alpha=0.5$ by almost one unit of $e/h$. (iv) The behavior of the spin Chern number of the VB, i.e., $C_s^{(2)}$ is shown over the full range of $\alpha$. Note that $C_s^{(2)}$ is calculated considering the entire FBZ, which is also consistent with the continuum model results given in Table \ref{Tab_Chern}(Appendix \ref{AppnA}).} 
\label{fig:Fig_SOM}
\end{figure}

In Figs. \ref{fig:Fig_SOM}(i) and \ref{fig:Fig_SOM}(ii), we show how the SROM evolves as $\mu$ is varied, respectively for $\alpha=0.3$
and $\alpha=0.7$. In both cases, $M_{\rm SR}^{\rm orb}$ varies linearly with   $\mu$ in the gap regions $\Delta_1$ or $\Delta_2$ or 
$\Delta_3$. The slopes of the SROM in $\Delta_1$, $\Delta_2$, and 
$\Delta_3$ calculated from Figs.\,\ref{fig:Fig_SOM}(i) and \ref{fig:Fig_SOM}(ii) are given in Table \ref{Table_Slope}.
\begin{table}[ht]
\centering
\setlength{\tabcolsep}{12pt}
\renewcommand{\arraystretch}{1.5}
\begin{tabular}{|c|c|c|c|}\hline

\multicolumn{1}{|c|}{Value of} &
\multicolumn{3}{|c|}{~~~~Slope of the SROM in~~~~} \\ \cline{2-4}

\multicolumn{1}{|c|}{$\alpha$} &
\multicolumn{1}{|c|}{$\Delta_1$} &
\multicolumn{1}{|c|}{$\Delta_2$} &
\multicolumn{1}{|c|}{$\Delta_3$}  \\ \hline
\ \ \ $0.3$ & $0.879$ & $1.043$ & $0.879$ \\
\hline
\ \ \ $0.7$ & $1.840$ & $1.497$ & $1.840$\\
\hline
\end{tabular}
\caption{The slope of the SROM versus the chemical potential (in units of $e/h$) in the forbidden gaps $\Delta_1$, $\Delta_2$, and 
$\Delta_3$ for two values of $\alpha$, namely $\alpha=0.3$ and $\alpha=0.7$.}
\label{Table_Slope}
\end{table}
The slope in $\Delta_1$ is same as that in $\Delta_3$ for a given value of 
$\alpha$.
The slope in $\Delta_2$ is higher than that in $\Delta_1$ by $0.164$ (in units of $e/h$) for $\alpha=0.3$. In contrast, we find the slope in $\Delta_2$ is lower than that in $\Delta_1$ by $0.343$ (in units of $e/h$) in the case of 
$\alpha=0.7$. Most importantly, we find that the slope either in $\Delta_1$ or in $\Delta_3$ for $\alpha=0.7$ differs from the same for $\alpha=0.3$ by almost one unit of $e/h$. This change in slope across $\alpha=0.5$ is essentially connected to the spin Chern number of the underlying system which we explain in the following.
As $\mu$ falls in $\Delta_1$, four bands, namely, spin-up and spin-down VBs in both valleys, i.e., 
$\varepsilon_{+,\uparrow}^2$, $\varepsilon_{+,\downarrow}^2$, 
$\varepsilon_{-,\uparrow}^2$, and $\varepsilon_{-,\downarrow}^2$ are occupied as understood from Figs. \ref{fig:Fig_VOM}(iii) and \ref{fig:Fig_VOM}(iv). In this case, 
Eq.\,(\ref{M_gap}) becomes
\begin{eqnarray}\label{Spin_slope1}
\frac{dM_{\rm SR}^{\rm orb}}{d\mu}=\frac{e}{h}C_s^{(2)},
\end{eqnarray}
where $C_s^{(2)}$ is the spin Chern number of the VB defined as
\begin{eqnarray}\label{Spin_Chern}
C_s^{(2)}=\frac{1}{2}\Big[C_{+,\uparrow}^{(2)}-C_{+,\downarrow}^{(2)}+
C_{-,\uparrow}^{(2)}-
C_{-,\downarrow}^{(2)}\Big].
\end{eqnarray}
Equation (\ref{Spin_slope1}) dictates that the slope of the SROM in $\Delta_1$ should be proportional to $C_s^{(2)}$. Our explicit calculations reveal that $C_s^{(2)}=1$ for $\alpha=0.3$ and $C_s^{(2)}=2$ for $\alpha=0.7$ (see Fig. \ref{fig:Fig_SOM}(iv)). The change in $C_s^{(2)}$ by one unit, therefore, mimics the change in the slope of the SROM in the band gap $\Delta_1$ by one unit of $e/h$.

As $\mu$ falls in $\Delta_2$, six bands such as $\varepsilon_{+,\uparrow}^2$, $\varepsilon_{+,\downarrow}^2$, 
$\varepsilon_{-,\uparrow}^2$, $\varepsilon_{-,\downarrow}^2$, $\varepsilon_{+,\downarrow}^1$, 
and $\varepsilon_{-,\uparrow}^1$ are occupied. Therefore,
from Eq.\,(\ref{M_gap}) we have
\begin{eqnarray}\label{Spin_slope2}
\frac{dM_{\rm SR}^{\rm orb}}{d\mu}=\frac{e}{h}\Big[C_s^{(2)}+\frac{1}{2}\big(C_{-,\uparrow}^{(1)}-C_{+,\downarrow}^{(1)}
\big)\Big].
\end{eqnarray}
Note that the quantity $C_{-,\uparrow}^{(1)}-C_{+,\downarrow}^{(1)}$ is precisely the valley Chern number of the VB, i.e., $C_v^{(2)}$. 
Equation\,(\ref{Spin_slope2}) becomes 
\begin{eqnarray}\label{Spin_slope23}
\frac{dM_{\rm SR}^{\rm orb}}{d\mu}=\frac{e}{h}\Big[C_s^{(2)}+\frac{1}{2}C_v^{(2)}\Big].
\end{eqnarray}
Therefore, the slope of the SROM in $\Delta_2$ is governed by the spin as well as the valley Chern number of the VB. 
We explicitly find that 
$C_v^{(2)}=0.330 (-0.684)$ for $\alpha=0.3 (0.7)$.  
From Eq.\,(\ref{Spin_slope23}) it is clear that the SROM should vary linearly in $\Delta_2$ with a slope which is higher (lower) than that in $\Delta_1$ in the case of $\alpha=0.3 (0.7)$. Eight bands, namely, $\varepsilon_{+,\uparrow}^2$, $\varepsilon_{+,\downarrow}^2$, 
$\varepsilon_{-,\uparrow}^2$, $\varepsilon_{-,\downarrow}^2$, $\varepsilon_{+,\downarrow}^1$,
$\varepsilon_{-,\uparrow}^1$, $\varepsilon_{+,\uparrow}^1$, and
$\varepsilon_{-,\downarrow}^1$ are occupied when $\mu$ is varied in the gap 
$\Delta_3$. In this case, we have 
\begin{eqnarray}\label{Spin_slope3}
\frac{dM_{\rm S}^{\rm orb}}{d\mu}=\frac{e}{h}\big[C_s^{(2)}+C_s^{(1)}\big].
\end{eqnarray}
We note that the spin Chern number of the FB, i.e., $C_s^{(1)}$ vanishes(see Table \ref{Tab_Chern}, Appendix \ref{AppnA}).
Equation (\ref{Spin_slope3}), therefore, tells us that the SROM would vary linearly with the chemical potential in $\Delta_3$ with the same slope as in $\Delta_1$.

We have also estimated the slope of the SROM in $\Delta_1$ over the entire range of $\alpha$ which is shown in Fig.\,\ref{fig:Fig_SOM}(iii). It is clear that the slope changes discontinuously by almost one unit of $e/h$ across 
$\alpha=0.5$. This abrupt change of the slope is indeed connected to the TPT from a quantum spin-Hall(QSH) phase with $C_s^{(2)}=1$ to another QSH phase with $C_s^{(2)}=2$ across $\alpha=0.5$ as depicted in 
Fig.\,\ref{fig:Fig_SOM}(iv).

The behavior of the OM at the TPT ($\alpha=0.5$) is worth mentioning here. As shown earlier, the Berry curvature $\Omega_{\eta,\sigma}^n(\bm k)$ and the OMM $m_{\eta,\sigma}^n(\bm k)$ are mostly concentrated near the valley i.e. $\bm k=0$. Both $\Omega_{\eta,\sigma}^n(\bm k)$ and $m_{\eta,\sigma}^n(\bm k)$ contain $(\varepsilon_{\eta,\sigma}^n-\varepsilon_{\eta,\sigma}^{n^\prime})$ in the denominator and diverge at the TPT where energy bands touch each other at ${\bm k}=0$, as evident from Fig.\,\ref{fig:Fig_BerryCPeak} and 
Fig.\,\ref{fig:Fig_Magmom_Peak}. However, these quantities remain well-defined at all other points in the Brillouin zone. Consequently, in the thermodynamic limit, at the TPT, the OM given in Eq.\,\eqref{Tot_OM} shows a divergent behavior and suggests the breakdown of the perturbation theory. However, the OM exists due to its experimental meaning.

\section{Circular Dichroism and Optical absorbance}\label{Sec5}
In this section, we are interested in the circular dichroism(CD) and low-frequency optical absorbance associated with the interband optical absorption.

\subsection{Circular Dichroism}
In general, the CD is a measure of the differential absorption of the left and the right circular polarized light. 
In the last section, we have seen that the OM could sense the TPT in a spin-orbit coupled $\alpha$-$T_3$ system. Since the OM and the CD are interconnected, it is tempting to ask whether the signatures of the TPT found in OM could also be found in the CD spectrum. In particular, we are interested in the valley-contrasted characteristics of the CD in two distinct QSH phases across $\alpha=0.5$.

We consider a circularly polarized light, incident vertically on the plane of the $\alpha$-$T_3$ lattice, which is described by the vector potential $\bm A=A(l\cos\omega t, \sin\omega t)$, where $A$ is the amplitude, $\omega$ is the angular frequency, and $l=\pm1$ is the polarization index. The light-matter interaction enters into the Hamiltonian through Peierls substitution $\bm k\rightarrow \bm k +e\bm A/\hbar$. As a result the Hamiltonian (Eq.\,(\ref{Ham_Valley})) becomes
\begin{eqnarray}
H_\sigma^\eta(\bm k, \bm A)=H_\sigma^\eta(\bm k)+\mathcal{P}_x^\eta A_x+\mathcal{P}_y^\eta A_y,
\end{eqnarray}
where $\mathcal{P}_x^\eta=\eta v_F S_x$ and 
$\mathcal{P}_y^\eta=-v_F S_y$ with
\begin{equation*}
S_x = \begin{pmatrix}
0 & \cos\phi & 0\\
\cos\phi & 0 & \sin\phi\\
0 & \sin\phi & 0
\end{pmatrix}
\end{equation*}
and
\begin{equation*}
S_y =\begin{pmatrix}
0 & -i\cos\phi & 0\\
i\cos\phi & 0 & -i\sin\phi\\
0 & i\sin\phi & 0
\end{pmatrix}.
\end{equation*}
Note that $\mathcal{P}_i^\eta$ ($i=x, y$) can also be identified as
\begin{eqnarray}
\mathcal{P}_i^\eta=\frac{1}{\hbar}\frac{\partial H_\sigma^\eta}{\partial k_i}.
\end{eqnarray}
The expressions of $\mathcal{P}_x^\eta$ and $\mathcal{P}_y^\eta$ do not contain any real spin index $\sigma$. Therefore, 
$\mathcal{P}_i^\eta$ is responsible for interband transitions between the states having same spin index at a given valley, thus defines a particular spin-valley selection rule. The transition matrix element between the initial state 
$\vert u_{\eta,\sigma}^n\ra$ and the final state $\vert u_{\eta,\sigma}^{n^\prime}\ra$ is given by
\begin{eqnarray}
P_\pm^{\eta,\sigma}(\bm k)=m_e\la u_{\eta,\sigma}^{n^\prime}(\bm k)\vert (\mathcal{P}_x^\eta\pm i\mathcal{P}_y^\eta)\vert u_{\eta,\sigma}^n(\bm k)\ra,
\end{eqnarray}
where $m_e$ is the free electron mass and the $+(-)$ sign corresponds to the right (left) circular polarized light. Note that
$\vert P_\pm^{\eta,\sigma}(\bm k)\vert^2$ represents the optical absorption in a definite valley $\eta$. The ${\bm k}$-resolved degree of optical polarization is given by
\begin{eqnarray}\label{Deg_pol}
\xi^{\eta,\sigma}(\bm k)=\frac{\vert P_+^{\eta,\sigma}(\bm k)\vert^2-\vert P_-^{\eta,\sigma}(\bm k)\vert^2}{\vert P_+^{\eta,\sigma}(\bm k)\vert^2+
 \vert P_-^{\eta,\sigma}(\bm k)\vert^2}.
\end{eqnarray}

\begin{figure}[h!]
\centering
\includegraphics[width=8.5cm, height=5cm]{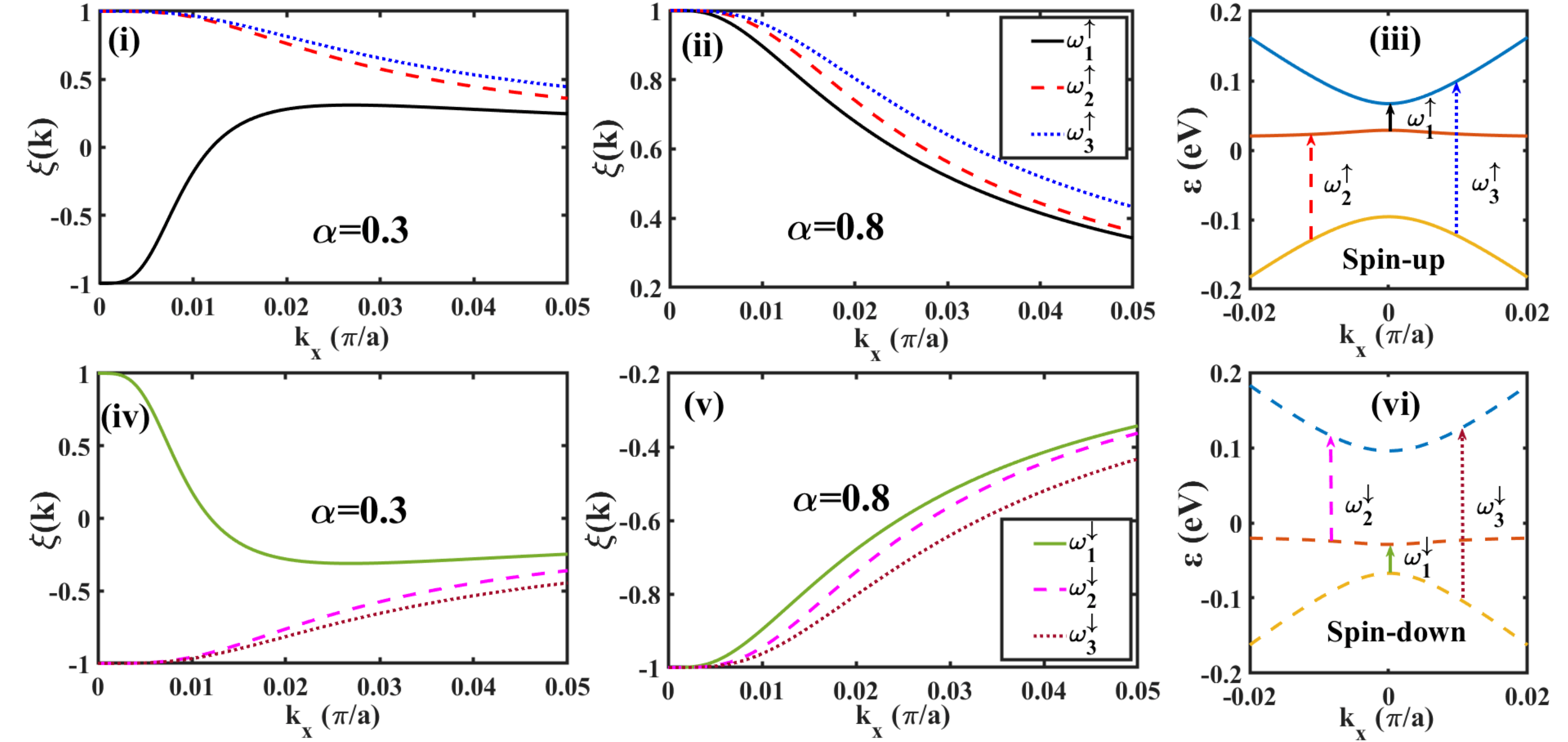}
\caption{The $\bm k$-resolved degree of optical polarization 
$\xi(\bm k)$ in the $K$ valley for the interband transitions 
$\omega_i^{\uparrow(\downarrow)}$($i=1,2,3$) in two distinct QSH phases corresponding to $\alpha=0.3$ and $\alpha=0.8$.} 
\label{fig:Fig_Dichro}
\end{figure}

The behavior of the $\bm k$-resolved optical polarization around the $K$ valley is shown in
Fig.\,\ref{fig:Fig_Dichro} for $\alpha=0.3$ and $\alpha=0.8$. 
Figs.\,\ref{fig:Fig_Dichro}(i)-(iii) correspond to spin-up bands while Figs.\,\ref{fig:Fig_Dichro}(iv)-(vi) are for spin-down bands. Possible interband transitions are depicted by up-arrows in 
Figs.\,\ref{fig:Fig_Dichro}(iii) and \ref{fig:Fig_Dichro}(vi) for both spin bands. For spin-up case, $\omega_1^\uparrow$, 
$\omega_2^\uparrow$, and 
$\omega_3^\uparrow$ denote the following interband transitions: FB($\varepsilon_{+,\uparrow}^1$)$\rightarrow$CB($\varepsilon_{+,\uparrow}^0$), VB($\varepsilon_{+,\uparrow}^2$)$\rightarrow$FB($\varepsilon_{+,\uparrow}^1$), and VB($\varepsilon_{+,\uparrow}^2$)$\rightarrow$CB($\varepsilon_{+,\uparrow}^0$), respectively. On the other hand, 
$\omega_1^\downarrow$, $\omega_2^\downarrow$, and 
$\omega_3^\downarrow$ correspond to 
VB($\varepsilon_{+,\downarrow}^2$)$\rightarrow$FB($\varepsilon_{+,\downarrow}^1$), FB($\varepsilon_{+,\downarrow}^1$)$\rightarrow$CB($\varepsilon_{+,\downarrow}^0$), and VB($\varepsilon_{+,\downarrow}^2$)$\rightarrow$CB($\varepsilon_{+,\downarrow}^0$) interband transitions, respectively, for spin down case. 

We first discuss the interband transitions between spin-up bands. We note that the optical polarizations are perfect for the band-edge excitations($k=0$), where the exact spin-valley optical selection rule is obeyed.
$\xi(\bm k)$ diminishes away from $k=0$. For both 
$\omega_2^{\uparrow}$- and $\omega_3^{\uparrow}$-transitions,
$\xi(\bm k)$ starts from $+1$ at $k=0$ and decreases away from the valley. This behavior of $\xi(\bm k)$ weakly depends on $\alpha$ as evident from Figs.\,\ref{fig:Fig_Dichro}(i) and \ref{fig:Fig_Dichro}(ii). However, 
$\xi(\bm k)=-1$ at $k=0$ for $\omega_1^\uparrow$-transition when $\alpha=0.3$. Its behavior drastically changes when $\alpha=0.8$. Moreover, it becomes $+1$ at $k=0$ which could be considered as a possible signal of the TPT.  For spin-down case the features of 
$\xi(\bm k)$ are exactly opposite to that of spin-up case.

\begin{figure}[h!]
\centering
\includegraphics[width=8cm, height=5.5cm]{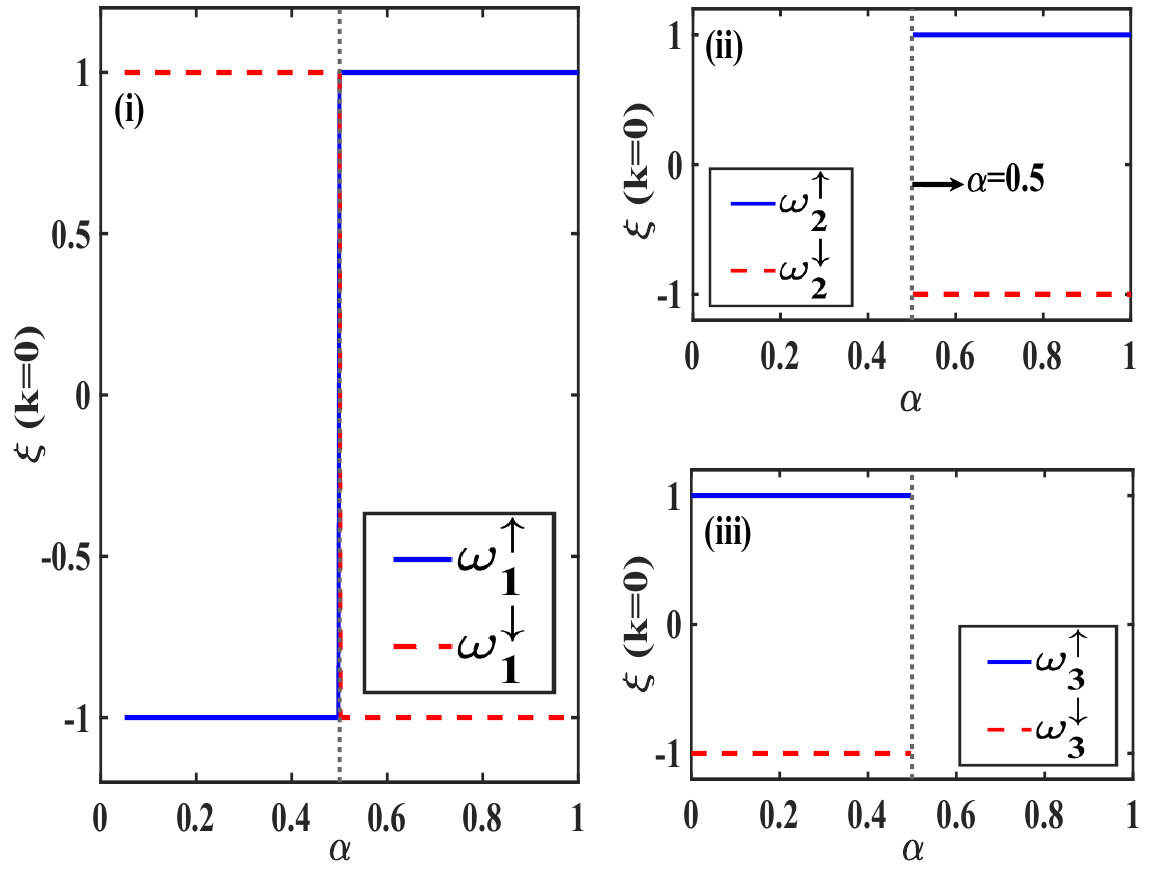}
\caption{The $\bm k$-resolved optical polarization at $k=0$ i.e. $\xi(\bm k=0)$ for various interband transitions as a function of $\alpha$.}
\label{fig:Fig_Dichro_alpha}
\end{figure}

In Fig.\,\ref{fig:Fig_Dichro_alpha}, we plot $\xi(\bm k=0)$ as function of $\alpha$ for all interband transitions. For 
$\omega_1^\uparrow\,(\omega_1^\downarrow)$-transition, $\xi(\bm k=0)$ changes from $-1\,(+1)$ to $+1\,(-1)$  across $\alpha=0.5$ as shown in
Fig.\,\ref{fig:Fig_Dichro_alpha}(i). This means that, at the $K$ valley, a left\,(right) circular polarized light is responsible for 
the $\omega_1^\uparrow\,(\omega_1^\downarrow)$-transition in QSH phase corresponding to $\alpha<0.5$ while a right\,(left) circular polarized light causes the same transition in the QSH phase corresponding to $\alpha>0.5$. We also note that $\omega_2^\uparrow$- and 
$\omega_2^\downarrow$-transitions are not allowed for $\alpha<0.5$ whereas $\omega_3^\uparrow$- and $\omega_3^\downarrow$-transitions are forbidden for $\alpha>0.5$ as depicted in 
Fig.\,\ref{fig:Fig_Dichro_alpha}(ii) and Fig.\,\ref{fig:Fig_Dichro_alpha}(iii), respectively.

The above mentioned features of $\xi(\bm k=0)$ can be explained by analyzing the energy band structure at $k=0$. We first focus on the spin-up bands in the $K$ valley. The energy eigenvalues for $\alpha<0.5$ are given by
$\varepsilon_{+,\uparrow}^0(0)=\lambda(\cos\phi-\sin\phi)$, 
$\varepsilon_{+,\uparrow}^1(0)=\lambda\sin\phi$, and
$\varepsilon_{+,\uparrow}^2(0)=-\lambda\cos\phi$
with corresponding eigenstates as
$\vert u_{+,\uparrow}^0\rangle=(0~1~0)^\mathcal{T}$, 
$\vert u_{+,\uparrow}^1\rangle=(0~0~1)^\mathcal{T}$, and 
$\vert u_{+,\uparrow}^2\rangle=(1~0~0)^\mathcal{T}$,
respectively, where $\mathcal{T}$ denotes the transpose.
On the other hand, for $\alpha>0.5$, we have energy eigenvalues:
$\varepsilon_{+,\uparrow}^0(0)=\lambda\sin\phi$, $\varepsilon_{+,\uparrow}^1(0)=\lambda(\cos\phi-\sin\phi)$, and 
$\varepsilon_{+,\uparrow}^2(0)=-\lambda\cos\phi$,
and corresponding eigenstates:
$\vert u_{+,\uparrow}^0\rangle=(0~0~1)^\mathcal{T}$,
$\vert u_{+,\uparrow}^1\rangle=(0~1~0)^\mathcal{T}$, and
$\vert u_{+,\uparrow}^2\rangle=(1~0~0)^\mathcal{T}$, respectively.
Clearly the roles of $\varepsilon_{+,\uparrow}^0$ and $\varepsilon_{+,\uparrow}^1$ are interchanged (i.e. band inversion) across $\alpha=0.5$, a typical characteristics of the TPT.

For the $\omega_1^\uparrow$-transition, we find
\begin{eqnarray}\label{P_omg1}
 && P_{+}^{+,\uparrow}=2m_ev_F\sin\phi\,\Theta(\alpha-1/2),\nonumber\\
&& P_{-}^{+,\uparrow}=2m_ev_F\sin\phi\,\Theta(1/2-\alpha),
\end{eqnarray}
where $\Theta(x)$ is the Heaviside step function. Therefore, $P_{+}^{+,\uparrow}=0$ for $\alpha<0.5$ which causes $\xi(\bm k=0)=-1$. For $\alpha>0.5$, we have $\xi(\bm k=0)=+1$ because
$P_{-}^{+,\uparrow}=0$.

For $\omega_2^\uparrow$ and $\omega_3^\uparrow$-transitions, we have
\begin{eqnarray}\label{P_omg2}
 P_{+}^{+,\uparrow}=2m_ev_F\,\cos\phi\,\Theta(\alpha-1/2),~~~
 P_{-}^{+,\uparrow}=0,
\end{eqnarray}
and
\begin{eqnarray}\label{P_omg3}
 P_{+}^{+,\uparrow}=2m_ev_F\,\cos\phi\,\Theta(1/2-\alpha),~~~
 P_{-}^{+,\uparrow}=0,
\end{eqnarray}
respectively. 
Note that Eqs.\,(\ref{P_omg1})-(\ref{P_omg3}) clearly explain the behaviour of $\xi(\bm k=0)$ as depicted in 
Figs.\,\ref{fig:Fig_Dichro_alpha}(i)-\ref{fig:Fig_Dichro_alpha}(iii). The treament for the spin-down case would be similar to that for the spin-up case.

The CD vanishes in the cases of dice lattice ($\alpha=1$) and graphene ($\alpha=0$) as a consequence of the SI symmetry. This can be proved in the following way.
For $\alpha=1$, we find that the  $\omega_3$-transition between the VB and the CB  is forbidden because $P_{\pm}^{\eta,\sigma}(\bm k)=0$. However, the $\omega_1$(FB$\rightarrow$CB) and $\omega_2$(VB$\rightarrow$ FB) -transitions are allowed and they are equally probable because the FB resides midway between the CB and the VB. In either cases, we obtain
\begin{eqnarray}
P_\pm^{\eta,\sigma}=\frac{m_e\hbar^2 v_F^3 k}{\sqrt{2}\varepsilon_d^2}\Bigg[\frac{\sqrt{2}\varepsilon_d\lambda}{2\varepsilon_d^2-\lambda^2}\pm\sigma\Big(1+\frac{\lambda^2}{2\varepsilon_d^2-\lambda^2}\Big)\Bigg]k_\pm,
\end{eqnarray}
where $\varepsilon_d=\sqrt{\hbar^2v_F^2k^2+\lambda^2/2}$ and $k_\pm=k_x\pm ik_y$. It is noteworthy that $P_\pm^{\eta,\sigma}$ is independent of the valley index $\eta$.
Inserting $P_\pm^{\eta,\sigma}$ in Eq. (\ref{Deg_pol}), one can obtain 
\begin{eqnarray}
\xi^{\eta,\sigma}(\bm k)=\sigma\frac{\sqrt{2}\,\lambda\,\varepsilon_d}{\varepsilon_d^2+\frac{\lambda^2}{2}}.
\end{eqnarray}
The energy bands of the dice lattice are spin and valley degenerate because of the SI and the TR symmetries. In order to obtain the resultant $\xi(k)$, we need to sum 
$\xi^{\eta,\sigma}(\bm k)$ over the degenerate spin states that gives rise to vanishing CD.

For $\alpha=0$, only the $\omega_3$-transition is allowed because the energy spectrum of graphene does not have any FB. In this case we find
\begin{eqnarray}
\xi^{\eta,\sigma}(\bm k)=\sigma\frac{2\,\lambda\,\varepsilon_g}{\varepsilon_g^2+\lambda^2},
\end{eqnarray}
where $\varepsilon_g=\sqrt{\hbar^2v_F^2k^2+\lambda^2}$. Likewise the dice lattice, $\xi(\bm k)$ vanishes as a result of the SI symmetry.

\subsection{Optical Absorbance}
Here, we are interested in the behavior of low-frequency optical absorbance across the TPT. In the case of normal incidence, the interband optical absorbance of a two-dimensional system is given by [\onlinecite{Absorbance_1}]
\begin{eqnarray}\label{Dielec1}
\tiny{\mathcal{A}_\pm^{j\rightarrow i}(\omega)=\frac{8\pi^2\hbar\alpha_f}{m_e^2\omega S}
\sum_{\bm k}\vert P_\pm^{j\rightarrow i}(\bm k)\vert^2
\delta[\varepsilon_i(\bm k)-\varepsilon_j(\bm k)-\hbar\omega]},
\end{eqnarray}
where $\alpha_f$ is the fine structure constant. Here, we mainly consider the band-edge absorption because the optical selection rules are exact at $k=0$. Replacing 
$P_\pm^{j\rightarrow i}(\bm k)$ by $P_\pm^{j\rightarrow i} (0)$, Eq.\,(\ref{Dielec1}) becomes
\begin{eqnarray}\label{Dielec2}
\mathcal{A}_\pm^{j\rightarrow i}(\omega)\simeq\frac{8\pi^2\hbar\alpha_f}{m_e^2\omega S}\vert P_\pm^{j\rightarrow i}(0)\vert^2N_{\rm opt}^{j\rightarrow i}(\hbar\omega),
\end{eqnarray}
where the optical joint density of states(JDOS) $N_{\rm opt}^{j\rightarrow i}(\hbar\omega)$ is given by
\begin{eqnarray}
 N_{\rm opt}^{j\rightarrow i}(\hbar\omega)=
\sum_{\bm k}\delta[\varepsilon_i(\bm k)-\varepsilon_j(\bm k)-\hbar\omega].\nonumber
\end{eqnarray}

Expanding the energy spectrum in Eq.\,(\ref{Eigen_Energy}) around $k=0$, we find 
\begin{eqnarray}\label{Energy_approx}
 \varepsilon^{n}_{\eta,\sigma}(\bm k)\approx
 \varepsilon^{n}_{\eta,\sigma}(0)+\frac{\hbar^2k^2}{2\widehat{m}_{n}^{\eta,\sigma}},
\end{eqnarray}
where the expressions of $\varepsilon^{n}_{\eta,\sigma}(0)$ and $\widehat{m}_{n}^{\eta,\sigma}$ are given in Appendix B. It is straightforward to obtain 
$N_{\rm opt}^{j\to i}(\hbar\omega)$ as
\begin{eqnarray}
N_{\rm opt}^{j\to i}(\omega)=\frac{\widehat{m}_{ij}^{\eta,\sigma}S}{2\pi\hbar^2}\Theta[\hbar\omega-\delta\varepsilon_{ij}^{\eta,\sigma}],
\end{eqnarray}
where $\delta\varepsilon_{ij}^{\eta,\sigma}=\varepsilon^{i}_{\eta,\sigma}(0)-\varepsilon^{j}_{\eta,\sigma}(0)$ and
$\widehat{m}_{ij}^{\eta,\sigma}=\frac{\widehat{m}_{i}^{\eta,\sigma}\widehat{m}_{j}^{\eta,\sigma}}{\widehat{m}_{j}^{\eta,\sigma}-\widehat{m}_{i}^{\eta,\sigma}}.$

We are mainly interested in the optical absorbance in the $K$ valley
for $\omega_1^\uparrow$- and $\omega_1^\downarrow$-transitions because we have seen earlier that $\xi(\bm k)$ exhibits sign change across the 
TPT for the aforesaid interband optical transitions. Therefore,
$\varepsilon_i(\bm k)=\varepsilon_{+,\uparrow}^0$ and $\varepsilon_j(\bm k)=\varepsilon_{+,\uparrow}^1$ for the 
$\omega_1^\uparrow$-transition and $\varepsilon_i(\bm k)=\varepsilon_{+,\downarrow}^1$ and $\varepsilon_j(\bm k)=\varepsilon_{+,\downarrow}^2$ for the $\omega_1^\downarrow$-transition. The optical JDOS becomes
\begin{eqnarray}
N_{\rm opt}^{\omega_1^\uparrow}=\frac{\widehat{m}_{01}^{+,\uparrow}S}{2\pi\hbar^2}\Theta[\hbar\omega-\delta\varepsilon_{01}^{+,\uparrow}]
\end{eqnarray}
and
\begin{eqnarray}
N_{\rm opt}^{\omega_1^\downarrow}=\frac{\widehat{m}_{12}^{+,\downarrow}S}{2\pi\hbar^2}\Theta[\hbar\omega-\delta\varepsilon_{12}^{+,\downarrow}],
\end{eqnarray}
for $\omega_1^\uparrow$- and $\omega_1^\downarrow$-transition, respectively. 

\begin{figure}[h!]
\centering
\includegraphics[width=8cm, height=6cm]{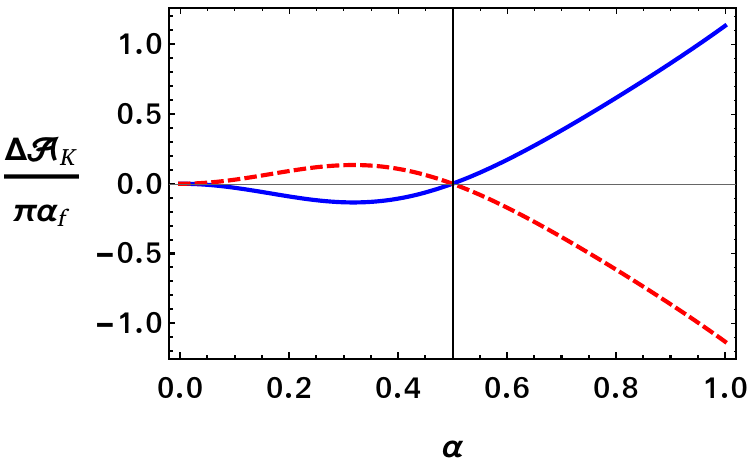}
\caption{The differential optical absorbance $\Delta\mathcal{A}_K$ as function of $\alpha$. Here, we set $\hbar\omega=5\lambda$. The solid\,(dashed) line represents the $\omega_1^\uparrow\,(\omega_1^\downarrow)$-transition. The solid vertical line is drawn at $\alpha=0.5$.}
\label{fig:Fig_Absorbance}
\end{figure}

Let us define the quantity $\Delta\mathcal{A}_K(\omega)=\mathcal{A}_+(\omega)-\mathcal{A}_-(\omega)$ which is a measure of differential absorbance of the right and left circular polarized light. For the 
$\omega_1^\uparrow$- and $\omega_1^\downarrow$-transitions, we obtain
\begin{eqnarray}
\tiny{\frac{\Delta\mathcal{A}_K^{\omega_1^\uparrow}}{\pi\alpha_f}=\frac{32\pi\hbar v_F^2\sin^2\phi}{\omega S}\Bigg[\Theta\Big(\alpha-\frac{1}{2}\Big)-\Theta\Big(\frac{1}{2}-\alpha\Big)\Bigg]N_{\rm opt}^{\omega_1^\uparrow}}.
\end{eqnarray}
and 
\begin{eqnarray}
\tiny{\frac{\Delta\mathcal{A}_K^{\omega_1^\downarrow}}{\pi\alpha_f}=\frac{32\pi\hbar v_F^2\sin^2\phi}{\omega S}\Bigg[\Theta\Big(\frac{1}{2}-\alpha\Big)-\Theta\Big(\alpha-\frac{1}{2}\Big)\Bigg]N_{\rm opt}^{\omega_1^\downarrow}}.
\end{eqnarray}
It is straightforward to show that $N_{\rm opt}^{\omega_1^\uparrow}=N_{\rm opt}^{\omega_1^\downarrow}$ because $\widehat{m}_{01}^{+,\uparrow}=\widehat{m}_{12}^{+,\downarrow}$ and $\delta\varepsilon_{01}^{+,\uparrow}=\delta\varepsilon_{12}^{+,\downarrow}$. Therefore, we have
$\Delta\mathcal{A}^{\omega_1^\downarrow}=-\Delta\mathcal{A}^{\omega_1^\uparrow}$.
The variation of the differential absorbance corresponding to $\omega_1^\uparrow$- and $\omega_2^\downarrow$-transitions with $\alpha$ are shown in Fig.\,\ref{fig:Fig_Absorbance}. It is clear that $\Delta\mathcal{A}_K$ changes its sign across $\alpha=0.5$.

\begin{figure}[h!]
\centering
\includegraphics[width=8cm, height=6cm]{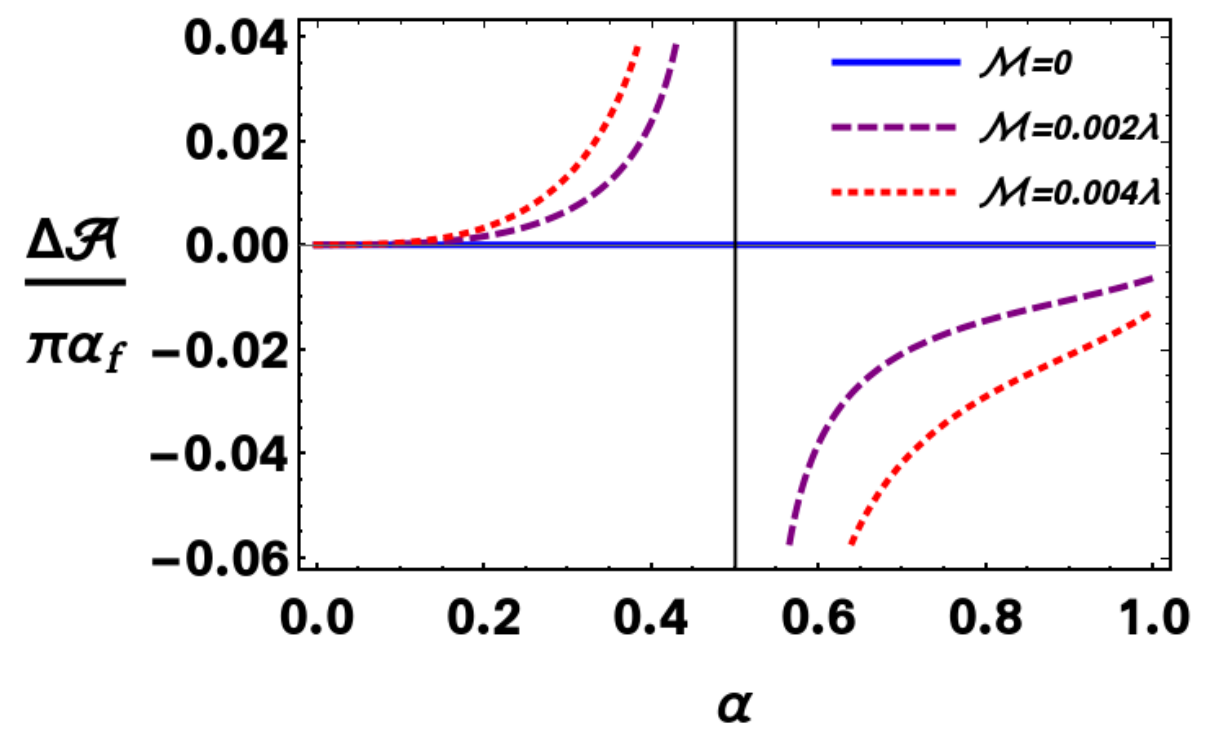}
\caption{The total differential optical absorbance as a function of $\alpha$ when the Fermi level is in the gap $\Delta_1$. The vertical line is drawn at
$\alpha=0.5$. We have chosen $\hbar\omega=5\lambda$.} 
\label{fig:Fig_ABsT}
\end{figure}

So far we have considered the contribution of the $K$ valley alone in the differential optical absorbance. Our main focus was to demonstrate how $\Delta \mathcal{A}_K$ changes sign across the TPT at $\alpha=0.5$. The summation over $\bm k$ in Eq. (\ref{Dielec1}) suggests that one should consider the contribution from the $K^\prime $ valley also. As a consequence of the TR symmetry, the total optical absobance vanishes, imposing a constraint on its experimental realization.
This obstacle can be overcome by studying higher order responses of the system such as nonlinear optical absorption and/or nonlinear conductivities. Even in TR symmetric systems, breaking of the SI symmetry may lead to nonlinear effects via finite Berry curvature dipole [\onlinecite{BCD}]. It is demonstrated that the Berry curvature dipole can detect a TPT in twisted double bilayer graphene [\onlinecite{BCD_TDBG}]. In our system, nonlinear effects can be found for $0<\alpha<1$ due to broken SI symmetry. Therefore, it would be interesting to see in near future whether such nonlinear effects can sense the TPT. Nevertheless, this issue can be addressed also by introducing the following term into the Hamiltonian that breaks the TR symmetry explicitly,
\begin{eqnarray}
H_\mathcal{M}=\mathcal{M}\sigma\begin{pmatrix}
1 & 0 & 0\\
0 & 0 & 0\\
0 & 0 & -1
\end{pmatrix}.\nonumber
\end{eqnarray}
This term can be interpreted as $A$-$C$ lattice staggered magnetization[\onlinecite{Spin_Hall_Phase}]. The behavior of the total differential optical absorbance (obtained by summing the contributions from both valleys) is shown in Fig. \ref{fig:Fig_ABsT} for $\mathcal{M}=0, 0.002\lambda$, and $0.004\lambda$.
$\Delta\mathcal{A}$ vanishes when $\mathcal{M}=0$ because of the TR symmetry. For $\mathcal{M}\neq0$, $\Delta\mathcal{A}$ is finite as TR symmetry is broken. Most importantly, it behaves differently in two topologically distinct QSH phases. In this way, the differential optical absorbance can be considered as an experimental marker to differentiate various topological phases. The technical details regarding $\mathcal{M}\neq0$ are provided in Appendix \ref{AppnC}.
The optical absorbance of graphene was determined using the optical reflectivity and transmission measurements [\onlinecite{Absorb_graphn}]. Similar techniques would also be helpful in realizing our results experimentally.

\section{Summary}\label{Sec6}
In summary, we have explored the topological features of the 
$\alpha$-$T_3$ system considering the SOI of Kane-Mele type. To be specific, we have studied the behavior of the Berry curvature, the OMM and the OM to probe the TPT across $\alpha=0.5$ between two distinct QSH phases. 
We have also discussed an alternative scheme for detecting the TPT through optical absorption as the underlying system exhibits a strong CD controlled by a particular spin-valley optical selection rule. We have reported that there is a discontinuous sign change in the Berry curvature as well as in the OMM across $\alpha=0.5$. This sign change has been argued as a possible signature of the TPT. For $0<\alpha<1$, the energy spectrum consists of three well separated forbidden gaps, $\Delta_1$, $\Delta_2$, and 
$\Delta_3$, introduced by the SOI. The signatures of the TPT across 
$\alpha=0.5$ have been captured by analyzing the behavior of the VROM and the SROM in those gaps, separately. The VROM increases linearly in $\Delta_1$, attains a plateau in $\Delta_2$, decreases linearly in $\Delta_3$ with the same slope as in $\Delta_1$ when the chemical potential scans the entire energy spectrum. The plateau region of the VROM in $\Delta_2$ offers to establish the relationship between the Chern number of the FB with that of the VB. The slope of the VROM versus the chemical potential in $\Delta_1$ changes its sign abruptly across $\alpha=0.5$, which is in tune with the corresponding behavior of the valley Chern number. On the other hand, the SROM varies linearly with the chemical potential in the forbidden gaps. The slope of the SROM in $\Delta_1$ is same as that in $\Delta_3$. Most strikingly, the slope of the SROM in $\Delta_1$  suddenly changes by almost one unit of $e/h$ across $\alpha=0.5$, a reminiscent of the corresponding change in the spin Chern number of the VB from $C_s^{(2)}=1$ to $C_s^{(2)}=2$. However, the slope of the SROM in $\Delta_2$ is higher(lower) than that in $\Delta_1$ for $\alpha<0.5$ ($\alpha>0.5$). It is controlled by both the spin and the valley Chern numbers of the VB. Finally, the $k$-resolved degree of optical polarization behaves differently for different interband transitions. Most strikingly, it changes sign across the TPT in the case of the interband transition between the bands that touch each other at $\alpha=0.5$. We obtain  analytical expressions of the low-frequency differential optical absorbance associated with different interband transitions in a definite valley, which also exhibit  sign change across the TPT. Finally, we discuss possible signatures of different QSH phases in the optical absorbance, that can be detected experimentally.  

\section*{ACKNOWLEDGMENTS}
S.V. sincerely acknowledges the financial support from the Institute for Basic Science (IBS) in the Republic of Korea through the project IBS-R024-D1.
T. B. would like to acknowledge the financial support provided
by the University of North Bengal through University Research Projects to pursue this work.

\appendix
\begin{widetext}
\section{Calculation of the Chern number}\label{AppnA}
Here, we provide the details of the calculation of the Chern number within  the continuum model. Using the expression of the Berry curvature given in 
Eq.\,(\ref{Berry_C}), one can calculate the Chern number $C_{\eta,\sigma}^{(n)}$ for a given valley and spin orientation from Eq.\,(\ref{Eq_Chern}). The calculated values of $C_{\eta,\sigma}^{(n)}$ for $\alpha=0.3$ and $\alpha=0.7$ are given in Table \ref{Tab_Chern}. For a given band $n$, spin and valley integrated Chern number vanishes i.e. 
$\sum_{\eta,\sigma} C_{\eta,\sigma}^{(n)}=0$. However, the spin Chern number of the VB, as defined in Eq.\,(\ref{Spin_Chern}), is $C_s^{(2)}=1$ for 
$\alpha=0.3$ and $C_s^{(2)}=2$ for $\alpha=0.7$. This result is consistent with the full FBZ calculations as shown in Fig.\,\ref{fig:Fig_SOM}(iv).

\begin{table}[h!]
\centering
\setlength{\tabcolsep}{14pt}
\renewcommand{\arraystretch}{2}
\begin{tabular}{|c|c|c|} \hline
$C_{\eta,\sigma}^{(n)}$ & $\alpha=0.3$ & $\alpha=0.7$ \\
\hline 
 $C_{+,\uparrow}^{(2)}$ & 0.581 &0.827\\ 
\hline
$C_{+,\downarrow}^{(2)}$ &-0.416 &-1.169\\
\hline
$C_{-,\uparrow}^{(2)}$ & 0.416 & 1.169\\
\hline
$C_{-,\downarrow}^{(2)}$ & -0.581 & -0.827\\
\hline
$C_{+,\uparrow}^{(1)}$ & -0.165 & 0.342\\
\hline
$C_{+,\downarrow}^{(1)}$ & -0.165 & 0.342\\
\hline
$C_{-,\uparrow}^{(1)}$ & 0.165 & -0.342\\
\hline
$C_{-,\downarrow}^{(1)}$ & 0.165 & -0.342\\
\hline
\end{tabular}
\caption{The Chern number $C_{\eta,\sigma}^n$ for a particular valley and spin orientation, calculated from Eq.\,(\ref{Eq_Chern}), for $\alpha=0.3$ and $\alpha=0.7$.}
\label{Tab_Chern}
\end{table}

\section{Approximate Energy spectrum close to the Dirac Point}\label{AppnB}
Here, we provide a detail calculation of the energy band structure very close to the Dirac point. We mainly expand the exact energy spectrum given in 
Eq.\,(\ref{Eigen_Energy}) around 
$k=0$. Retaing terms up to $O(k^2)$, we find
\begin{eqnarray}
\sqrt{\frac{-p}{3}}\approx \frac{1}{\sqrt{3}}\Big(\lambda_\phi+\frac{\hbar^2v_F^2}{2\lambda_\phi}k^2\Big)~~~~\mbox{and}~~~~
\frac{3q}{2p}\sqrt{\frac{-3}{p}}\approx \chi_\phi^{\eta,\sigma}+\Lambda_\phi^{\eta,\sigma}k^2,
\end{eqnarray}
where
\begin{eqnarray}
\lambda_\phi=\lambda\sqrt{1-\sin\phi\cos\phi},~~
 \chi_\phi^{\eta,\sigma}=\eta\sigma\frac{3\sqrt{3}}{2}\frac{(\sin\phi-\cos\phi)\sin\phi\cos\phi}{(1-\sin\phi\cos\phi)^{\frac{3}{2}}},~~\mbox{and}~~~
 \Lambda_\phi^{\eta,\sigma}=-
 \frac{\hbar^2v_F^2\sin^2(\phi+\pi/4)}{\lambda_\phi^2}\chi_\phi^{\eta,\sigma}.
\end{eqnarray}

Using the standard expansion $\arccos(a+bx^2)=\arccos(a)-bx^2/\sqrt{1-a^2}+O(x^3)$, the energy spectrum near $k=0$ can be obtained as
\begin{eqnarray}
 \varepsilon^{n}_{\eta,\sigma}(\bm k)\approx
 \varepsilon^{n}_{\eta,\sigma}(0)+\frac{\hbar^2k^2}{2\widehat{m}_{n}^{\eta,\sigma}},
\end{eqnarray}
where 
\begin{eqnarray}
\varepsilon^{n}_{\eta,\sigma}(0)=\frac{2\lambda_\phi}{\sqrt{3}}\cos\Bigg[\frac{\arccos(\chi_\phi^{\eta,\sigma})-2\pi n}{3}\Bigg]
\end{eqnarray}
and 
\begin{eqnarray}\label{massM0}
\frac{1}{\widehat{m}_{n}^{\eta,\sigma}}=\frac{2v_F^2}{\sqrt{3}\lambda_\phi}\Bigg\{\cos\Bigg[\frac{\arccos(\chi_\phi^{\eta,\sigma})-2\pi n}{3}\Bigg]-\frac{2}{3}\frac{\chi_{\phi}^{\eta,\sigma}}{\sqrt{1-(\chi_\phi^{\eta,\sigma})^2}}\sin^2(\phi+\pi/4)\sin\Bigg[\frac{\arccos(\chi_\phi^{\eta,\sigma})-2\pi n}{3}\Bigg]\Bigg\}.
\end{eqnarray}

\section{Calculation of the differential optical absorbance for $\mathcal{M}\neq0$}\label{AppnC}
Here, we provide the calculation of the differential optical absorbance in the presence of a staggered magnetization explicitly.
The TR symmetry is broken explicitly by considering the following $A$-$C$ lattice staggered magnetization term in the Hamiltonian [\onlinecite{Spin_Hall_Phase}]: 
\begin{eqnarray}
H_\mathcal{M}=\mathcal{M}\sigma\begin{pmatrix}
1 & 0 & 0\\
0 & 0 & 0\\
0 & 0 & -1
\end{pmatrix}.\nonumber
\end{eqnarray}

The total Hamiltonian becomes
\begin{eqnarray}
H_\sigma^\eta(\bm{k},\mathcal{M})&=&H_\sigma^\eta(\bm{k})+H_\mathcal{M}\nonumber\\
&=&
\begin{pmatrix}
\mathcal{M}\sigma-\lambda\eta\sigma\cos\phi & f_{\bm{k}} \cos\phi & 0\\
f_{\bm{k}}^* \cos\phi & \lambda\eta\sigma(\cos\phi-\sin\phi) & f_{\bm{k}}\sin\phi\\
0 & f_{\bm{k}}^*\sin\phi & -\mathcal{M}\sigma+\lambda\eta\sigma\sin\phi
\end{pmatrix}.\nonumber
\end{eqnarray}

The exact diagonalization of $H_\sigma^\eta(\bm k,\mathcal{M})$ gives rise to the following energy spectrum:
\begin{eqnarray}\label{Energy_Mag}
\varepsilon_{\eta,\sigma}^n({\bm k},\mathcal{M})=2\sqrt{\frac{-P_\mathcal{M}}{3}} \cos \Bigg[\frac{1}{3}\arccos\Bigg(\frac{3Q_\mathcal{M}}{2P_\mathcal{M}}\sqrt{\frac{-3}{P_\mathcal{M}}}\Bigg)-\frac{2\pi n}{3}\Bigg],
\end{eqnarray}
where
\begin{eqnarray}
&& P_\mathcal{M}=\lambda^2\sin\phi\cos\phi-(\lambda^2+\hbar^2v_F^2k^2)-\mathcal{M}^2-\mathcal{M}\lambda\eta(\sin\phi+\cos\phi)\nonumber\\
&& Q_\mathcal{M}=(\cos\phi-\sin\phi)\Big[\lambda\eta\sigma\Big\{(\lambda^2+\hbar^2 v_F^2 k^2)\sin\phi\cos\phi+\mathcal{M}^2\Big\}-\mathcal{M}\sigma(\lambda^2+\hbar^2v_F^2k^2)(\cos\phi+\sin\phi)\Big].\nonumber
\end{eqnarray}
Note that Eq.\,(\ref{Energy_Mag}) reduces to Eq.\,(\ref{Eigen_Energy}) in the limit $\mathcal{M}\to 0$.

It is explicitly shown [\onlinecite{Spin_Hall_Phase}] that the introduction of the staggered magnetization $\mathcal{M}$ into the Hamiltonian gives rise to a number of topological phases such as QSH, quatum anomalous Hall, quantum spin quantum anomalous (QSQAH), and antiferromagnetic phases characterized by $(C, C_s)$, where $C$ is the Chern number and $C_s$ is the spin Chern number. The system undergoes a TPT across $\alpha=0.5$ from a pure QSH phase with $(0, 1)$ to another pure QSH phase with $(0, 2)$ when $\mathcal{M}=0$ [\onlinecite{Spin_Hall_Phase}]. However, these two QSH phases are separated by a QSQAH phase with $(1, 3/2)$ for a finite value of $\mathcal{M}$. We restrict ourselve to consider
$\mathcal{M}\ll\lambda$ so that these QSH phases remain pure. For a given value of $\mathcal{M}$, we have two values of $\alpha$ that define the phase boundaries between these phases,
\begin{eqnarray}
\alpha_1=\frac{2\lambda^2-\sqrt{5\mathcal{M}^2\lambda^2-\mathcal{M}^4}}{4\lambda^2-\mathcal{M}^2}~~~\mbox{and}~~~\alpha_2=\frac{2\lambda^2+\sqrt{5\mathcal{M}^2\lambda^2-\mathcal{M}^4}}{4\lambda^2-\mathcal{M}^2}.\nonumber
\end{eqnarray} 
$\alpha_1$ and $\alpha_2$ merges to $\alpha=0.5$ in the limit 
$\mathcal{M}\to 0$. The QSH phase with $C_s=1$($C_s=2$) exists for $\alpha<\alpha_1$($\alpha>\alpha_2$) when $\mathcal{M}\neq 0$.

Following the same technique, described in Appendix \ref{AppnB}, we obtain the energy spectrum in the vicinity of the Dirac point ($\bm k=0$) 
\begin{eqnarray}
\varepsilon^{n}_{\eta,\sigma}(\bm k,\mathcal{M})\approx
\varepsilon^{n}_{\eta,\sigma}(0,\mathcal{M})+\frac{\hbar^2k^2}{2\widetilde{m}_{n}^{\eta,\sigma}(\mathcal{M})},
\end{eqnarray}
where 
\begin{eqnarray}
\varepsilon^{n}_{\eta,\sigma}(0,\mathcal{M})=\frac{2\lambda_{\phi,\mathcal{M}}}{\sqrt{3}}\cos\Bigg[\frac{\arccos(\chi_{\phi,\mathcal{M}}^{\eta,\sigma})-2\pi n}{3}\Bigg]
\end{eqnarray}
and 
\begin{eqnarray}\label{Imas_mag}
\frac{1}{\widetilde{m}_{n}^{\eta,\sigma}(\mathcal{M})}=\frac{2v_F^2}{\sqrt{3}\lambda_{\phi,\mathcal{M}}}\Bigg\{\cos\Bigg[\frac{\arccos(\chi_{\phi,\mathcal{M}}^{\eta,\sigma})-2\pi n}{3}\Bigg]+\frac{2}{3\hbar^2v_F^2}\frac{\lambda_{\phi,\mathcal{M}}^2\,\Xi_{\phi,\mathcal{M}}^{\eta,\sigma}}{\sqrt{1-(\chi_{\phi,\mathcal{M}}^{\eta,\sigma})^2}}\sin\Bigg[\frac{\arccos(\chi_{\phi,\mathcal{M}}^{\eta,\sigma})-2\pi n}{3}\Bigg]\Bigg\}.
\end{eqnarray}

Here,
\begin{eqnarray}
\lambda_{\phi,\mathcal{M}}=\sqrt{\lambda^2(1-\sin\phi\cos\phi)+\mathcal{M}^2+\mathcal{M}\lambda\eta(\sin\phi+\cos\phi)},
\end{eqnarray}

\begin{eqnarray}
\chi_{\phi,\mathcal{M}}^{\eta,\sigma}=\frac{3\sqrt{3}}{2}\frac{(\sin\phi-\cos\phi)}{\lambda_{\phi,\mathcal{M}}^3}\Bigg[\Big\{\lambda\eta\sigma\sin\phi\cos\phi-\mathcal{M}\sigma(\sin\phi+\cos\phi)\Big\}\lambda^2 + {\mathcal{M}}^2\lambda\eta\sigma\Bigg],
\end{eqnarray}
and 
\begin{eqnarray}
\frac{\Xi_{\phi,\mathcal{M}}^{\eta,\sigma}}{\hbar^2 v_F^2}=\frac{3\sqrt{3}}{2}\frac{(\sin\phi-\cos\phi)}{\lambda_{\phi,\mathcal{M}}^3}\Bigg[\Big\{\lambda\eta\sigma\sin\phi\cos\phi-\mathcal{M}\sigma(\sin\phi+\cos\phi)\Big\}\Big(1-\frac{3\lambda^2}{2\lambda_{\phi,\mathcal{M}}^2}\Big)-\frac{3{\mathcal{M}}^2\lambda\eta\sigma}{2\lambda_{\phi,\mathcal{M}}^2}\Bigg].
\end{eqnarray}
It is straightforward to verify that Eq.\,(\ref{Imas_mag}) reduces to Eq.\,(\ref{massM0}) in the limit $\mathcal{M}\to 0$.

The expression of the interband optical absorbance, derived in Eq.\,(\ref{Dielec2}), is given by
\begin{eqnarray}
\mathcal{A}_\pm^{j\to i}(\omega)\simeq\frac{8\pi^2\hbar\alpha_f}{m_e^2\omega S}\vert P_\pm^{j\to i}(0)\vert^2N_{\rm opt}^{j\to i}(\hbar\omega).
\end{eqnarray}
Here, the optical joint density of states $N_{\rm opt}^{j\to i}(\hbar\omega)$ is
\begin{eqnarray}
N_{\rm opt}^{j\to i}(\omega)=\frac{\widetilde{m}_{ij}^{\eta,\sigma}S}{2\pi\hbar^2}\Theta[\hbar\omega-\delta\varepsilon_{ij}^{\eta,\sigma}],
\end{eqnarray}
where $\delta\varepsilon_{ij}^{\eta,\sigma}=\varepsilon^{i}_{\eta,\sigma}(0,\mathcal{M})-\varepsilon^{j}_{\eta,\sigma}(0,\mathcal{M})$ and
$\widetilde{m}_{ij}^{\eta,\sigma}=\frac{\widetilde{m}_{i}^{\eta,\sigma}\widetilde{m}_{j}^{\eta,\sigma}}{\widetilde{m}_{j}^{\eta,\sigma}-\widetilde{m}_{i}^{\eta,\sigma}}.$

The quantity $\Delta\mathcal{A}^{j\to i}(\omega)=\mathcal{A}_+^{j\to i}(\omega)-\mathcal{A}_-^{j\to i}(\omega)$ measures the differential absorbance of the right and left circular polarized light. Let us consider the situation, depicted in Fig.\,\ref{fig:Fig_FermiL}, where the Fermi level $E_F$ lies in the gap $\Delta_1$ i.e. spin-up and spin-down VBs in both valleys are occupied. In the low-frequency regime, the optical transition between spin-down VB ($\varepsilon_{+,\downarrow}^2$) and spin-down FB ($\varepsilon_{+,\downarrow}^1$) in the $K$ valley and that between  between spin-up VB ($\varepsilon_{-,\uparrow}^2$) and spin-up FB ($\varepsilon_{-,\uparrow}^1$) in the $K^\prime$ valley are possible. These transitions are denoted by red and green arrows, respectively.

\begin{figure}[h!]
\centering
\includegraphics[width=10cm, height=8cm]{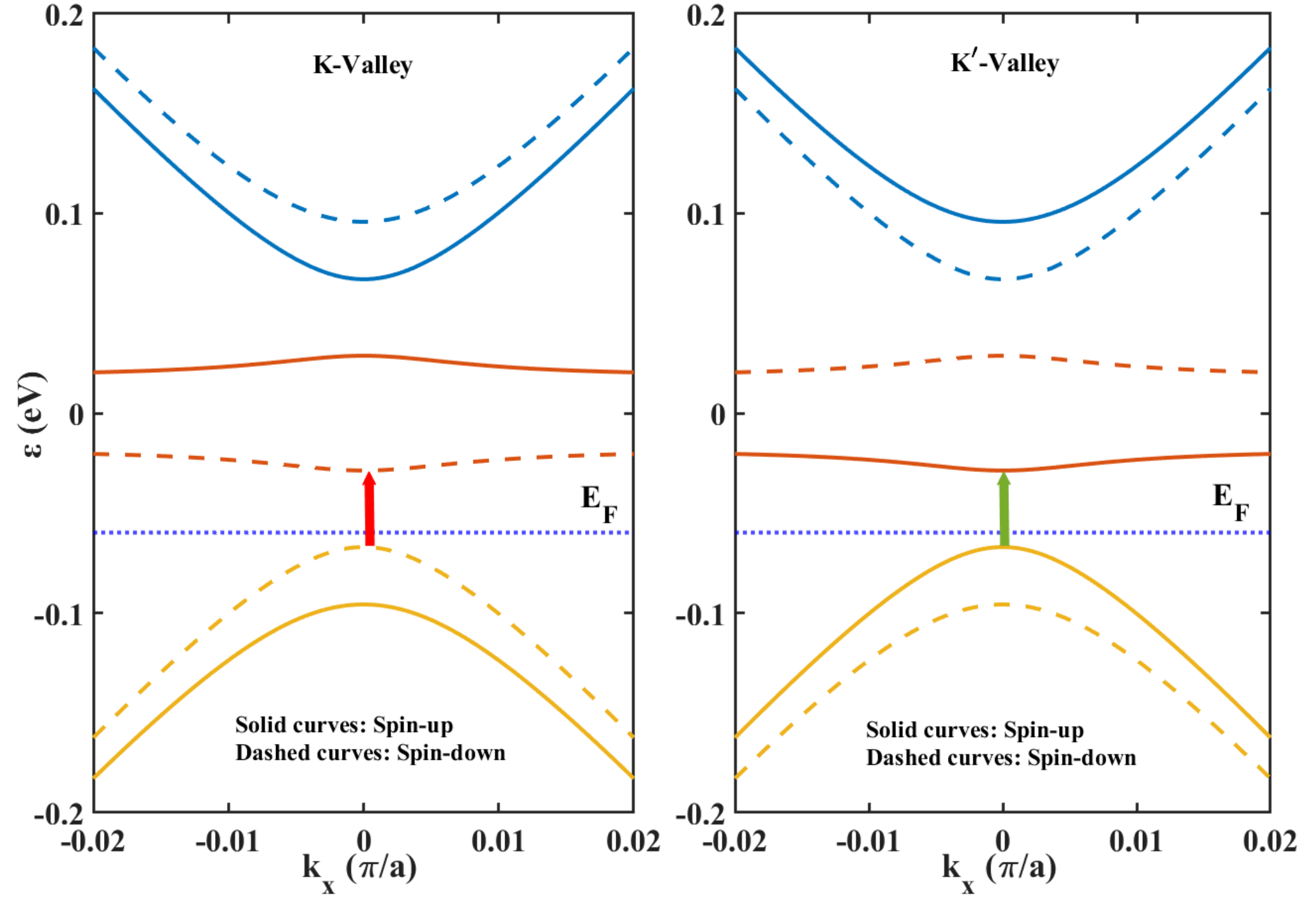}
\caption{Spin-split energy bands of a spin-orbit coupled $\alpha$-$T_3$ system. The blue dotted line represents the Fermi level. In the low-frequency regime, possible optical transitions in $K$ and $K^\prime$ valley are denoted by the red and green arrows, respectively.} 
\label{fig:Fig_FermiL}
\end{figure}

For the above mentioned optical transition in the $K$ valley, the transition matrix elements are given by
\begin{eqnarray}
&&P_+^{K,\downarrow}=2m_ev_F\sin\phi\,\Theta(\alpha_1-\alpha)\nonumber\\
&&P_-^{K,\downarrow}=2m_ev_F\sin\phi\,\Theta(\alpha-\alpha_2).\nonumber
\end{eqnarray}

For the $K^\prime$ valley, the transition matrix elements are 
\begin{eqnarray}
&&P_+^{K^\prime,\uparrow}=-2m_ev_F\sin\phi\,\Theta(\alpha-\alpha_2)\nonumber\\
&&P_-^{K^\prime,\uparrow}=-2m_ev_F\sin\phi\,\Theta(\alpha_1-\alpha).\nonumber
\end{eqnarray}
The effect of $\mathcal{M}$ on the transition matrix elements is clearly seen in the arguments of the $\Theta$-functions.

The differential absorbance associated with the optical transitions in the $K$ valley and the $K^\prime$ valley are given by
\begin{eqnarray}
\Delta\mathcal{A}_K^\downarrow(\omega)=\frac{4\pi\alpha_f}{\hbar\omega}\frac{(\vert P_+^{K,\downarrow}\vert^2-\vert P_-^{K,\downarrow}\vert^2)}{m_e^2}\widetilde{m}_{12}^{K,\downarrow}\,\Theta(\hbar\omega-\delta\varepsilon_{12}^{K,\downarrow})
\end{eqnarray} 
and 
\begin{eqnarray}
\Delta\mathcal{A}_{K^\prime}^\uparrow(\omega)=\frac{4\pi\alpha_f}{\hbar\omega}\frac{(\vert P_+^{K^\prime,\uparrow}\vert^2-\vert P_-^{K^\prime,\uparrow}\vert^2)}{m_e^2}\widetilde{m}_{12}^{K^\prime,\uparrow}\,\Theta(\hbar\omega-\delta\varepsilon_{12}^{K^\prime,\uparrow}),
\end{eqnarray}  
respectively. The total optical absorbation is given by $\Delta\mathcal{A}(\omega)=\Delta\mathcal{A}_K^\downarrow(\omega)+\Delta\mathcal{A}_{K^\prime}^\uparrow(\omega)$.

In Fig. \ref{fig:Fig_MKKp}, we show how $\widetilde{m}_{12}^{K,\downarrow}$ (upper panel) and $\widetilde{m}_{12}^{K^\prime,\uparrow}$ (lower panel) behave with $\alpha$ when $\mathcal{M}=0, 0.002\lambda,$ and $0.004\lambda$. We note that $\widetilde{m}_{12}^{K,\downarrow}$ coincides with $\widetilde{m}_{12}^{K^\prime,\uparrow}$ for $\mathcal{M}=0$, as a consequence of the TR symmetry. However, we have
$\widetilde{m}_{12}^{K,\downarrow}\neq\widetilde{m}_{12}^{K^\prime,\uparrow}$ when $\mathcal{M}\neq0$. In a similar spirit, one can also have $\delta\varepsilon_{12}^{K,\downarrow}\neq \delta\varepsilon_{12}^{K^\prime,\uparrow}$ for $\mathcal{M}\neq0$.

\begin{figure}[h!]
\centering
\includegraphics[width=8cm, height=9cm]{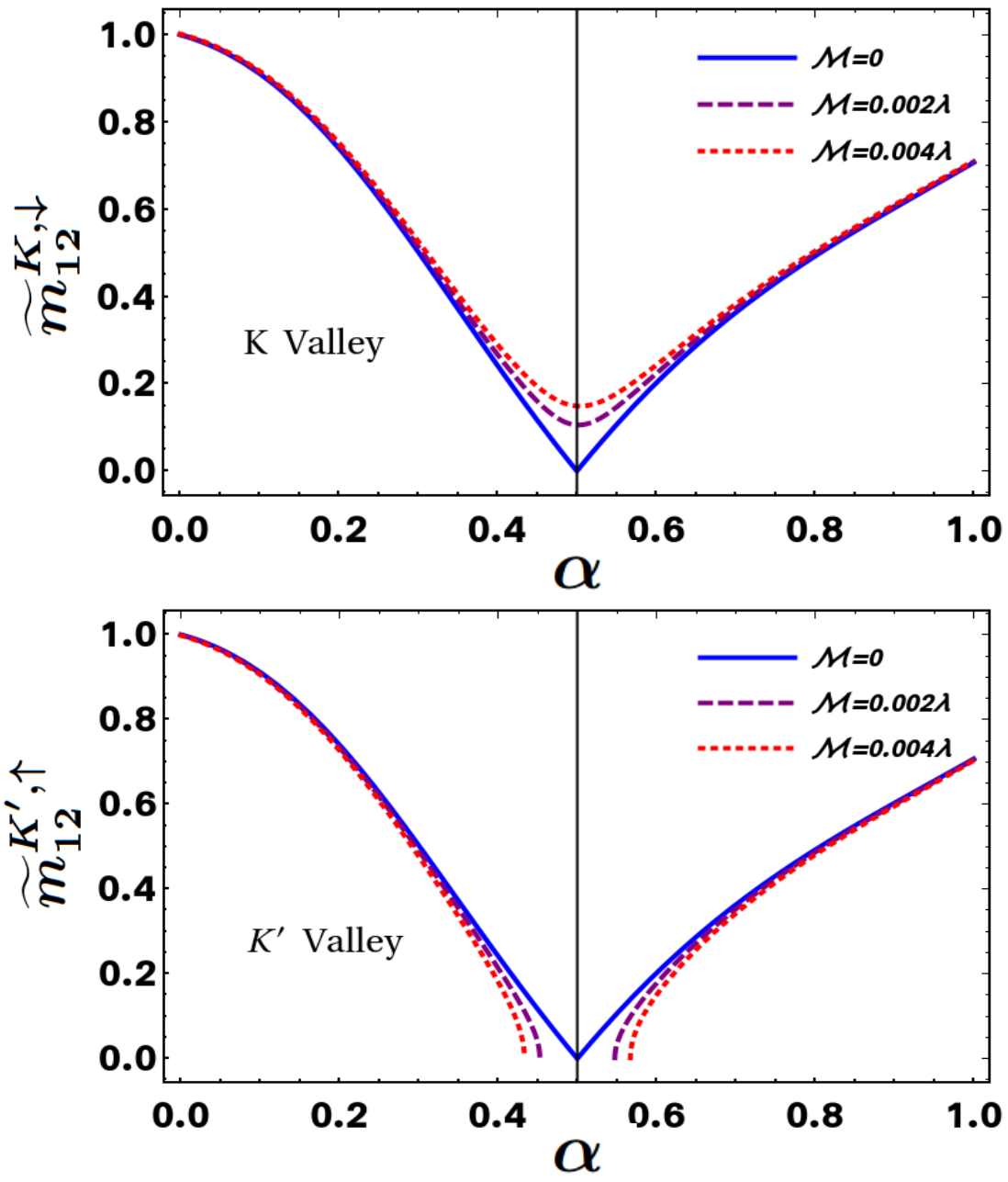}
\caption{The behavior of $\widetilde{m}_{12}^{K,\downarrow}$ (upper panel) and $\widetilde{m}_{12}^{K^\prime,\uparrow}$ (lower panel) as a function of $\alpha$ for different values of $\mathcal{M}$. $\widetilde{m}_{12}^{K,\downarrow}$ coincides with $\widetilde{m}_{12}^{K^\prime,\uparrow}$ for $\mathcal{M}=0$. However, $\widetilde{m}_{12}^{K,\downarrow}\neq\widetilde{m}_{12}^{K^\prime,\uparrow}$ when $\mathcal{M}\neq0$.} 
\label{fig:Fig_MKKp}
\end{figure}
\end{widetext}

\end{document}